\documentclass[aps,prfluids,print,superscriptaddress]{revtex4-1}

\usepackage{amsmath,amsfonts,amsthm,amssymb,bm} 
\usepackage{subfigure}
\usepackage{xcolor}
\usepackage{float}
\usepackage{comment}
\usepackage{todonotes}
\usepackage{natbib}
\usepackage{siunitx}

\newcommand{\latexe}{{\LaTeX\kern.125em2\lower.5ex\hbox{$\varepsilon$}}}

\chardef\bslash=`\\

\makeatletter
\def\square{\RIfM@\bgroup\else$\bgroup\aftergroup$\fi\vcenter{\hrule\hbox{\vrule\@height.6em\kern.6em\vrule}\hrule}\egroup}
\makeatother

\newcommand{\overbar}[1]{\mkern 1.5mu\overline{\mkern-1.5mu#1\mkern-1.5mu}\mkern 1.5mu}

\def\la{\langle}
\def\ra{\rangle}

\def\v2{\overline{v^2}}
\def\v2f{\overline{v^2}f}

\def\cG{\mathcal{G}}
\def\cA{\mathcal{A}}
\def\cM{\mathcal{M}}
\def\cD{\mathcal{D}}
\def\cF{\mathcal{F}}
\def\cI{\mathcal{I}}
\def\bkappa{{\bm\kappa}}

\definecolor{dy}{rgb}{0.6,0.6,0.1}
\definecolor{dg}{rgb}{0,0.6,0}

\begin{document}


\title{Resolution-induced anisotropy in LES}


\author{Sigfried W. Haering}
\affiliation{The Oden Institute for Computational Engineering and Science,\\ The University of Texas at Austin}
\author{Myoungkyu Lee}
\affiliation{Sandia National Laboratories, Livermore}
\author{Robert D. Moser}
\affiliation{Department of Mechanical Engineering,\\ The Oden Institute for Computational Engineering and Science, The University of Texas at Austin}


\date{\today}

\begin{abstract}
Large eddy simulation (LES) of turbulence in complex geometries and
domains is often conducted with high aspect ratio resolution cells of
varying shapes and orientations.  The effects of such anisotropic
resolution are often simplified or neglected in subgrid model
formulation. Here, we examine resolution induced anisotropy and
demonstrate that, even for isotropic turbulence, anisotropic
resolution induces mild resolved Reynolds stress anisotropy and
significant anisotropy in second-order resolved velocity gradient
statistics. In large eddy simulations of homogeneous isotropic
turbulence with anisotropic resolution, it is shown that commonly used
subgrid models, including those that consider resolution anisotropy in
their formulation, perform poorly. The one exception is the
anisotropic minimum dissipation model proposed by Rozema \emph{et al.}
(\emph{Phys. of Fluids} {\bf 27}, 085107, 2015). A simple new model is
presented here that is formulated with an anisotropic eddy diffusivity
that depends explicitly on the anisotropy of the resolution. It also
performs well, and is remarkable because unlike other LES subgrid
models, the eddy diffusivity only depends on statistical
characteristics of the turbulence (in this case the dissipation rate),
not on fluctuating quantities. In other subgrid modeling formulations,
such as the dynamic procedure, limiting flow dependence to statistical
quantities in this way could have advantages.
\end{abstract}


\maketitle


\section{Introduction}
\label{sec:intro}

The expansion of available computing power has led to an increasing
reliance on numerical simulation of complex systems in engineering,
science and decision making. With this increased reliance comes a
demand for modeling accuracy and reliability in general and
specifically in turbulent fluid flows. The increased resolution
enabled by advances in computing hardware, and increased numerical
accuracy that has arisen from advances in numerical algorithms has
improved the reliability of computational models of turbulent flows,
but improvements in turbulence models have lagged behind.  It has long
been expected that large eddy simulation (LES) would address the need
for improved modeling fidelity in engineering flows.  However,
numerous challenges remain before LES can become a robust tool capable
of reliable predictions of complex turbulent flows for use in research
and development.  Since the advent of the dynamic modeling
approach \cite{germ:1991,lilly:1992}, wall modeling has been
considered the greatest impediment to reliable LES, and this has been
the focus of much LES
research \cite{piom:2002,lars:2015,bose:2018}. While this is certainly
a critical issue, there are also other challenges. One such is
considered here.
 
In practical flows of engineering interest, the combination of high
Reynolds number, complex geometry and limited computational resources
often dictates discretization with relatively coarse, highly
anisotropic and spatially varying resolution. In such cases, the
common assumptions of isotropic unresolved turbulence in equilibrium
with the large scales and homogeneous filtering will generally be
violated. In this work, we focus in particular on the consequences of
anisotropic numerical resolution and the associated anisotropic
definition of the large (and small) scales. We will refer to this
simply as anisotropic resolution.

Though the issue of anisotropic resolution has been acknowledged in
many SGS models \cite{scott:1993,vrem:2004,roze:2015}, the specific
issue has not been examined in detail.  In this work, we examine the
implications of anisotropic resolution in LES and propose a simple
modeling treatment which has potential to be integrated into existing
models.  Before continuing, we briefly review existing resolution
anisotropy treatments in the literature.  Some of the models mentioned
below are evaluated in Sec.~\ref{sec:eval}.

In most cases, resolution anisotropy has been acknowledged through the
definition of a scalar resolved length scale in terms of anisotropic
resolution parameters. This resolved scale is then used in the
formulation of subgrid models, such as the Smagorinsky model.  
However, scalar measures discard all the information about 
resolution anisotropy.  For instance, consider the commonly used 
cube-root of a cell volume given by
$\Delta_{eq}=(\Delta_1\Delta_2\Delta_3)^{1/3}$, where $\Delta_\alpha$ 
are the resolution scales in each direction in an orthogonal grid.  This 
length scale will favor the smallest dimension of the grid and will thus 
provide an unresolvable model scale in coarse directions.  Such a 
simplification results in LES turbulence that is essentially under resolved 
in these directions and causing spectral energy pile ups at the resolved 
scale.  Conversely, scalar resolution measures based on the cell diagonal 
favor the largest dimension of the grid resulting in LES turbulence that is 
smoother in fine directions than could be resolved. This is essentially a 
waste of resolution.  Further, without explicit filtering corresponding 
to the cell diagonal scale, the spectral energy distribution will be effected 
with finer grid scales present.  In short, the LES turbulence is inconsistent with the 
with anisotropic filtering of real turbulence.

In an attempt to alleviate under-resolution in coarse directions,
Scotti \emph{et.al.} \cite{scott:1993} introduced a scalar correction
to $\Delta_{eq}$ in the standard Smagorinsky model, based on the ratio
of the refined to most coarse grid dimensions in an attempt to ensure the correct
total dissipation.  While an improvement over the basic Smagorinsky
model because it reduces artifacts of under-resolution in the course
directions, it is at the cost of even worse
under-utilization of available resolution in the fine directions. It
appears that these simple model forms preclude LES that produce
spectra consistent with anisotropic resolution.

The Vreman model \cite{vrem:2004}, which was primarily designed to
ensure that the eddy viscosity vanishes in laminar regions, was the
first to directly consider resolution anisotropy in its formulation.
In this model, the eddy viscosity scales with the square root of the
second invariant of the velocity gradient tensor with gradient 
components weighted by the corresponding grid length scale where the 
cells are assumed to be aligned with the global coordinate system. 
The magnitude of the resulting eddy viscosity is reduced in regions 
where high gradients are aligned with fine resolution directions and 
vice-versa.  However, in the end, resolution anisotropy information is 
again discarded in favor of maintaining a scalar eddy viscosity.  As 
shown in Sec.~\ref{sec:eval}, the result with anisotropic resolution is 
similar to basic Smagorinsky.

Rozema \emph{et al.} have more recently extended minimal dissipation
models \cite{vers:2011} to account for grid anisotropy (AMD), without making a scalar
filter width approximation like that described above \cite{roze:2015}.
Their model is motivated by the Ponc\`aire inequality applied to
anisotropic (rectilinear) grid cells, and is formulated to ensure that
the eddy viscosity is sufficient to dissipate energy at the estimated
rate of small-scale energy production. Here the anisotropy of the
resolution enters into the estimate of the small-scale production.
The AMD model performs quite
well with anisotropic resolution and it is also examined in some detail
in Sec.~\ref{sec:eval}.

Here, we pursue an evaluation of the impact of resolution anisotropy
on large eddy simulation models by simulating isotropic turbulence
with anisotropic resolution. The subgrid models described above that
consider resolution anisotropy, along with Smagorinsky, are evaluated.
Furthermore, we develop and evaluate a simple anisotropic tensor eddy
viscosity model. The model is particularly simple in that the eddy
viscosity does not fluctuate, though it does depend on statistical
properties of the turbulence being simulated, in this case the
dissipation.

In the remainder of the paper, the characteristics of
resolution-induced anisotropies in LES are discussed in
Sec.~\ref{sec:aniso}; our simple anisotropic subgrid model is
introduced in Sec.~\ref{sec:M43}; and, the performance of subgrid
models in isotropic turbulence simulated with anisotropic resolution
is explored in Sec.~\ref{sec:eval}. Finally, discussion and conclusions
are offered in Sec.~\ref{sec:conclusion}. 

\section{Resolution-induced anisotropies}
\label{sec:aniso}

In a large eddy simulation with anisotropic resolution, the anisotropy
of the resolution is expected to produce anisotropy of the resolved
and the subgrid Reynolds stresses. Of course, in homogeneous isotropic
turbulence (HIT), the Reynolds stress is dynamically
insignificant. However, in an LES of HIT with anisotropic resolution,
the resolved and subgrid Reynolds stress can be anisotropic, though
their sum is still isotropic and homogeneous. To examine how the
resolved and subgrid Reynolds stress anisotropy depend on resolution
anisotropy, consider an idealized infinite Reynolds number isotropic
turbulence with a $|\bkappa|^{-5/3}$ inertial range spectrum starting
at minimum wavenumber $\kappa_m$.  The Reynolds stress anisotropy can
be determined by integrating the inertial range energy spectrum over
an anisotropic domain of resolved wavenumbers $\cD$. That is, the
resolved and unresolved Reynolds stresses
($\la{\bar{u}_i\bar{u}_j}\ra$ and $\la{u'_iu'_j}\ra$, respectively)
are given by
\begin{equation}
\la{\bar{u}_i\bar{u}_j}\ra=\frac{C_k\varepsilon^{2/3}}{4\pi}\int_\cD|\bkappa|^{-11/3}\bigg{(}\delta_{ij}-\frac{\kappa_i\kappa_j}{|\bkappa|^2}\bigg{)}{}d\boldsymbol{\kappa}
\label{uu_res}
\end{equation}
and
\begin{equation}
\la{u'_iu'_j}\ra=\frac{C_k\varepsilon^{2/3}}{4\pi}\int_{\tilde\cD}|\bkappa|^{-11/3}\bigg{(}\delta_{ij}-\frac{\kappa_i\kappa_j}{|\bkappa|^2}\bigg{)}{}d\boldsymbol{\kappa}
\label{uu_unres}
\end{equation}
where $\tilde\cD$ is the domain of unresolved wavenumbers.

In this
paper, we consider two anisotropic definitions of the resolved
wavenumber domain. The first is an ellipsoidal wavenumber domain, with
major axes determined by the cutoff wavenumber
$\kappa_{c\alpha}=\pi/\Delta_\alpha$ in each direction, where
$\Delta_\alpha$ is the Nyquist grid spacing in the $\alpha$
direction. An ellipsoidal wavenumber domain is analogouos to the
spherical wavenumber cut-off commonly used in LES of isotropic
turbulence. Arguably ellipsoidally filtered turbulence is the most
meaningful target of an isotropic turbulence LES with anisotropic resolution. The
ellipsoidal resolved domain $\cD^e$ and the associated domain of
unresolved wavenumbers $\tilde\cD^e$ are given by 
\begin{align}
  \cD^e
  &= \left\{\bkappa \;\left|\; \sum_{\alpha=1}^3 \kappa_\alpha^2>\kappa^2_m\quad
  \mathrm{and}\quad
  \sum_{\alpha=1}^3 \kappa_\alpha^2\Delta^2_\alpha < \pi^2\right.\right\},\nonumber\\
  \tilde\cD^e &= \left\{\bkappa \;\left|\; \sum_{\alpha=1}^3 \kappa_\alpha^2\Delta^2_\alpha > \pi^2\right.\right\},
  \label{DefinecDe}
\end{align}
where $\kappa_m$ is the minimum wavenumber characterizing the largest
represented scale.  Note that non-tensor
indices are indicated by Greek letters; here because neither
$\kappa_{c\alpha}$ nor $\Delta_\alpha$ are index representations of
vectors.  No summation will be implied on repeated Greek
indices. The other anisotropic wavenumber domain considered here is
consistent with a
Cartesian tensor product representation of the LES solution in
physical space, where the resolution is different in each of the Cartesian
basis directions. This Cartesian domain $\cD^c$ is defined by
\begin{align}
  \cD^c &= \{\bkappa \mid \kappa_m < |\kappa_\alpha| < \kappa_{c\alpha};\; \alpha=1,2,3\},\nonumber\\
  \tilde\cD^c &= \{\bkappa \mid |\kappa_\alpha| > \kappa_{c\alpha};\;
  \alpha=1,2,3\}.
  \label{DefinecDc}
\end{align}
This definition of $\cD^c$ is anisotropic both because of the
Cartesian representation and because the $\Delta_\alpha$ are not in
general equal. The latter is most significant, and is of primary
interest here.

 \begin{figure}[tp]
 \begin{center}
 \subfigure[Unresolved]{\includegraphics
    [width=0.45\linewidth,bb= 48 0 466 420,clip]{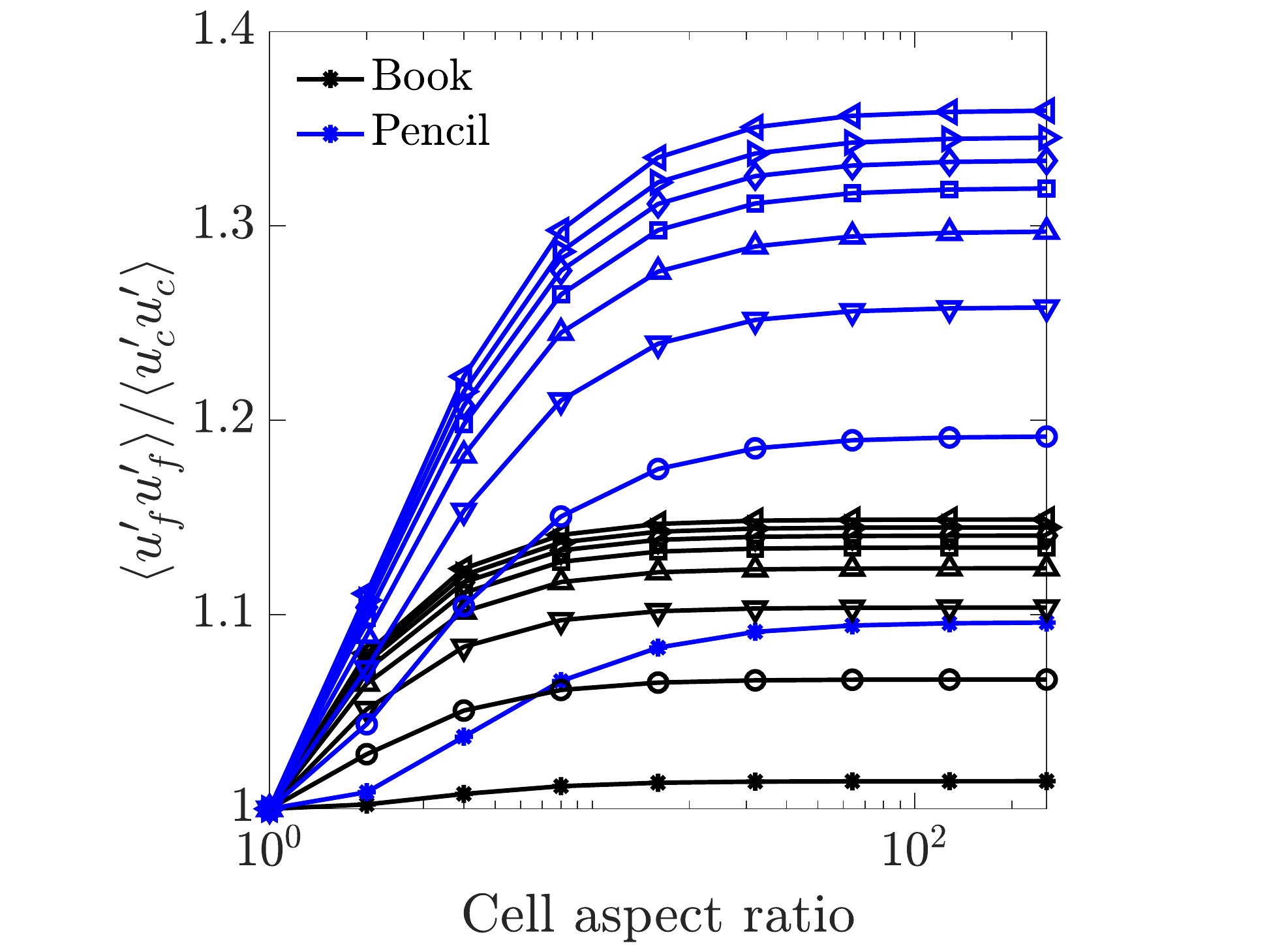} \label{Unresolved}}
 \subfigure[Resolved]{\includegraphics
    [width=0.45\linewidth,bb= 48 0 466 420,clip]{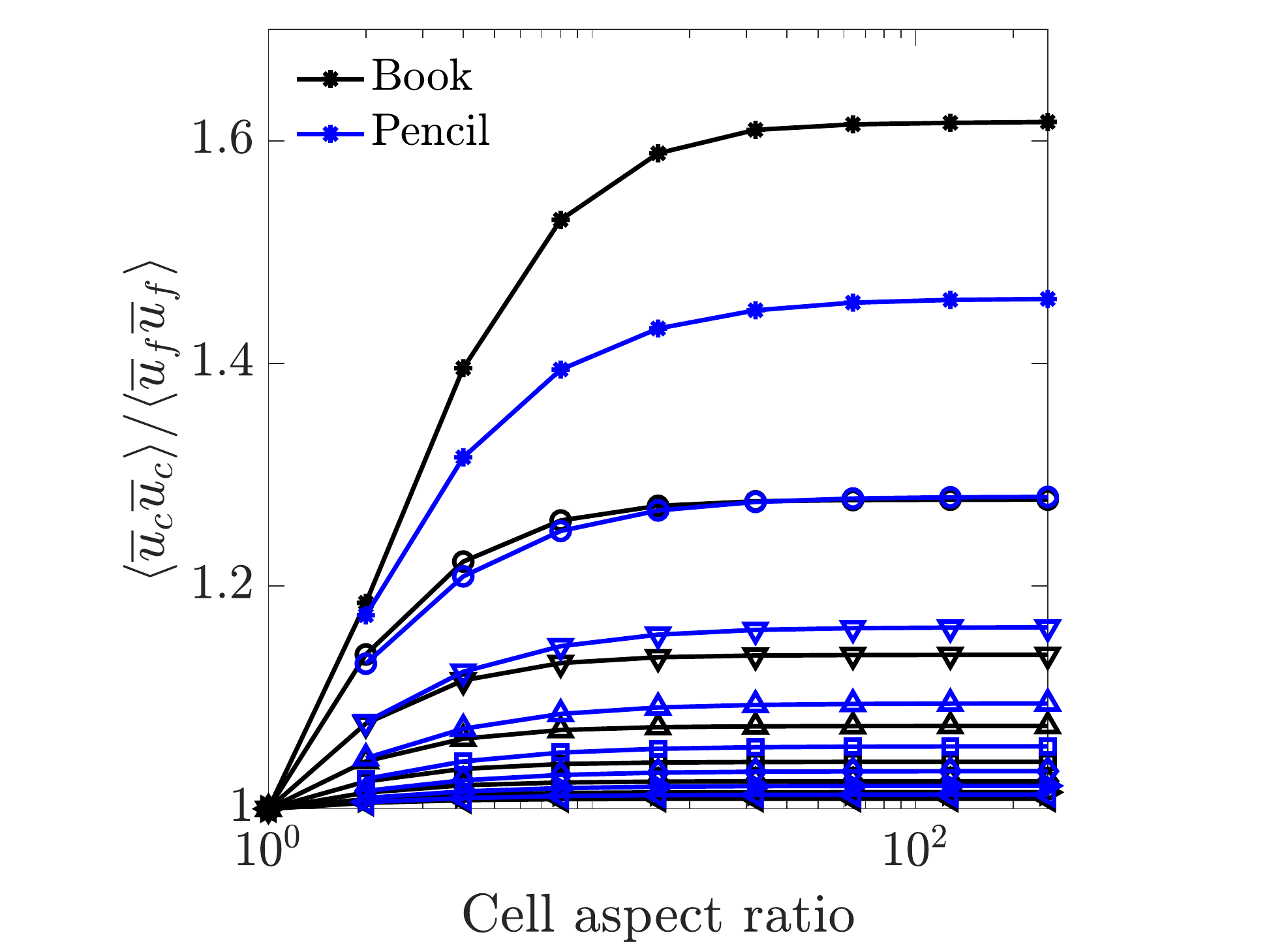} \label{Resolved}} \\
 \end{center}
 \caption{Resolution-induced anisotropy in the unresolved
   (a) and resolved (b) scale contributions to the Reynolds stress for an inertial range energy
   spectrum, determined as half the trace of (\ref{uu_unres}) and
   (\ref{uu_res}) based on the wavenumber domains $\cD^e$ and $\tilde\cD^e$ given by (\ref{DefinecDe}).
   The value of $k_{cc}/k_m$ is 2
   *, 4 $\circ$, 8 $\triangledown$, 16 $\vartriangle$, 32 $\square$, 64
   $\lozenge$, 128
   $\vartriangleright$, 256 $\vartriangleleft$.
   Subscript ``c'' indicates the coarse direction while ``f'' indicates the fine direction.}
 \label{e_aniso}
 \end{figure}

The $\Delta_\alpha$ define an effective resolution cell, essentially a
grid cell when using a grid-based numerical representation. We will
use the ``resolution cell'' nomenclature to refer to a volume of the
domain with dimensions defined by the $\Delta_\alpha$, even when using
numerical methods for which there is no such grid (e.g. spectral
methods). Throughout this work, only  
cell shapes with at least two of the $\Delta_\alpha$ equal
are considered.  These limiting cases are labeled ``book''
cells when the repeated $\Delta_\alpha$ is largest and ``pencil'' cell
when the small $\Delta_\alpha$ is repeated. 
All other cell shapes would exhibit behavior intermediate between these cases.  Book type 
cells are particularly common, as they arise naturally to resolve boundary layers. 
Pencil type cells are typically used more sparingly but are often employed in critical regions such 
as near stagnation and separation points on 2D bodies.

For isotropic turbulence with anisotropic LES resolution, the
unresolved Reynolds stress is weakly anisotropic
(Fig. \ref{e_aniso}a), with anisotropy saturating at cell aspect
ratios of about 32. The saturated level of anisotropy increases with
the ratio $\kappa_{cc}/\kappa_m$ but since this is occurring as the
resolution of the LES is being refined so that the contribution of the
unresolved turbulence to the Reynolds stress is becoming negligible,
this anisotropy becomes increasing irrelevant to the model \cite{moser:2000}.  Pencil
cells produce significantly more unresolved Reynolds stress anisotropy
than book cells.  Resolved stress
anisotropy (Fig. \ref{e_aniso}b) similarly saturates at the same cell
aspect ratio and, as expected, is only significant when the smallest
$\kappa_{cc}/\kappa_m$ approaches one.  When this ratio 16 or greater,
the resolved anisotropy is negligible.




If the contribution of unresolved scales to the Reynolds stress were
the only relevant function of a SGS model, these mild anisotropies would indicate 
the effects of resolution anisotropy could be neglected. However,
as is well known, the primary impact of the unresolved turbulence in well-resolved LES is
as a sink of resolved energy. Nearly all subgrid models, which are
designed to represent this energy transfer, are formulated in terms of
the resolved velocity gradient tensor, so its resolution induced
anisotropy will be important. The energy transfer of a eddy viscosity
based subgrid model will necessarily be expressed in terms of second
order (or higher) moments of the velocity-gradient
tensor. Particularly, if a scalar eddy viscosity is uncorrelated with
the velocity gradients, the anisotropy of the energy transfer to the
small scales is determined directly for the anisotropy of the second
moment of the velocity gradient. Its anisotropy is examined
next.

The anisotropy of the second moment of the velocity gradient tensor
$\mathcal{G}_{ijkl}=\la{\partial_k\bar{u}_i\partial_l\bar{u}_j}\ra$
induced by anisotropic resolution in isotropic turbulence is
calculated by taking the gradient of the spectral energy density tensor
twice and again integrating over an anisotropic resolved wavenumber
domain $\cD$. Noting that
$\la{\bar{u}_i\partial_k\partial_l\bar{u}_j}\ra=\la{\partial_k\partial_l\bar{u}_i\bar{u}_j}\ra=0$,
one obtains
\begin{equation}
\mathcal{G}_{ijkl}
= \la{\partial_k\bar{u}_i\partial_l\bar{u}_j}\ra{} 
= \frac{C_k\varepsilon^{2/3}}{4\pi}\int_\cD\kappa_k\kappa_l|\bkappa|^{-11/3}\bigg{(}\delta_{ij}-\frac{\kappa_i\kappa_j}{|\bkappa|^2}\bigg{)}{}d\vec{\kappa},
\label{Gijkl}
\end{equation}
where $\cD$ is the domain of resolved wavenumbers. 
The structure of $\cG_{ijkl}$ as given in (\ref{Gijkl}) implies that
it can be written in terms of a fourth-ranked tensor $\cA_{ijkl}$,
which is invariant to all permutations of its indices:
\begin{equation}
  \cG_{ijkl}=\frac{C_k\varepsilon^{2/3}}{4\pi}\big(\delta_{ij}\cA_{mmkl}-\cA_{ijkl}\big).
\end{equation}
When the resolved wavenumber domain $\cD$ is symmetric about three
mutually orthogonal planes through the origin, as is the case for the
domains defined in (\ref{DefinecDe}) and (\ref{DefinecDc}), and the basis in
which the tensors are expressed are normal to these symmetry planes, 
reflection symmetries require that the elements of
$\cG_{ijkl}$ are zero unless each index is equal to at least one
other index, with the same requirements applying to $\cA_{ijkl}$.
Under these conditions, there are only six distinct non-zero
components of $\cA$, which can be organized in a symmetric matrix $A$:
\begin{equation}
  \cA_{\alpha\alpha\beta\beta}=A_{\alpha\beta}=\int_\cD
  \kappa_\alpha\kappa_\alpha\kappa_\beta\kappa_\beta|\bkappa|^{-17/3}\,d\vec{\kappa}
  \label{Aijkl}
\end{equation}
where again there is no summation over repeated Greek indices.  In this
basis, the remaining components of $\cA$ are either zero or equal to
those defined in (\ref{Aijkl}), by virtue of the symmetry and
invariance properties described above. When the LES resolution is
isotropic (i.e. $\cD$ is spherically symmetric), each element of
$A$ has one of only two distinct values, one for the diagonal
elements, and one for the off-diagonal. 

\begin{figure}[tp]
 \begin{center}
 \subfigure[All repeating]{\includegraphics
    [width=0.3\linewidth,bb= 62 0 485 420,clip]{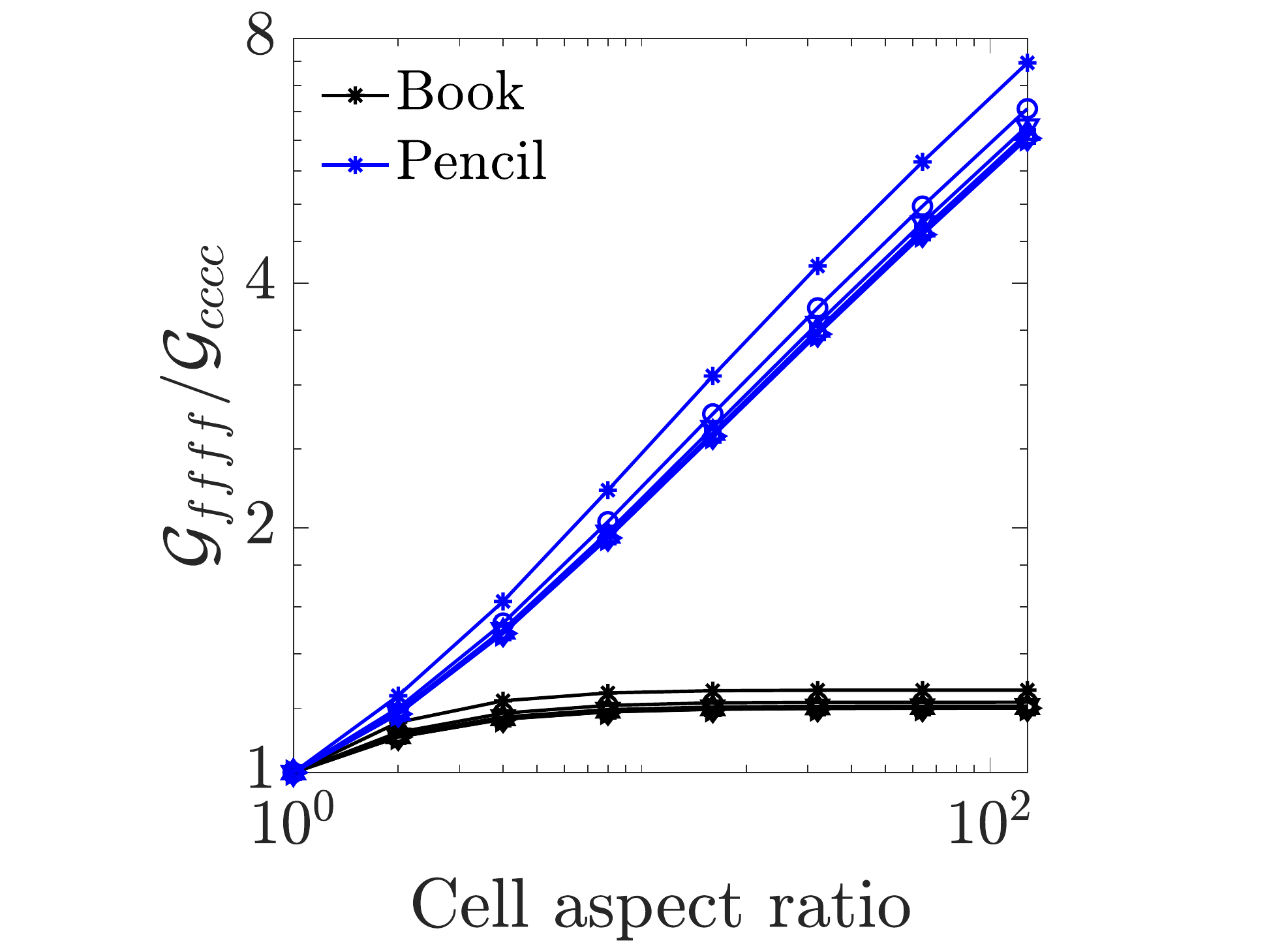}\label{Ga}}
 \subfigure[Fine component, coarse gradient]{\includegraphics
    [width=0.3\linewidth,bb= 62 0 485 420,clip]{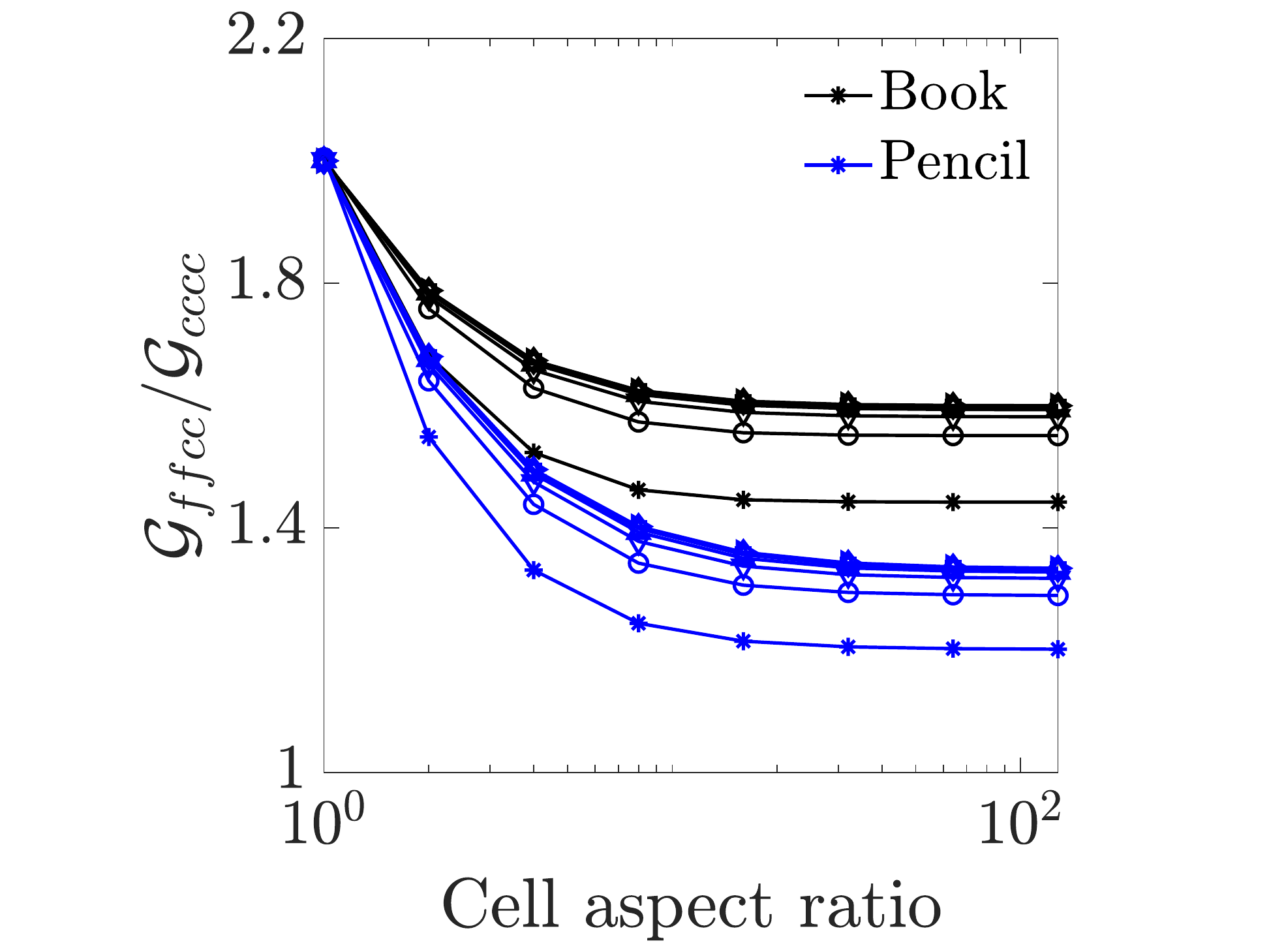}\label{Gb}}
 \subfigure[Coarse component, fine gradient]{\includegraphics
    [width=0.3\linewidth,bb= 62 0 485 420,clip]{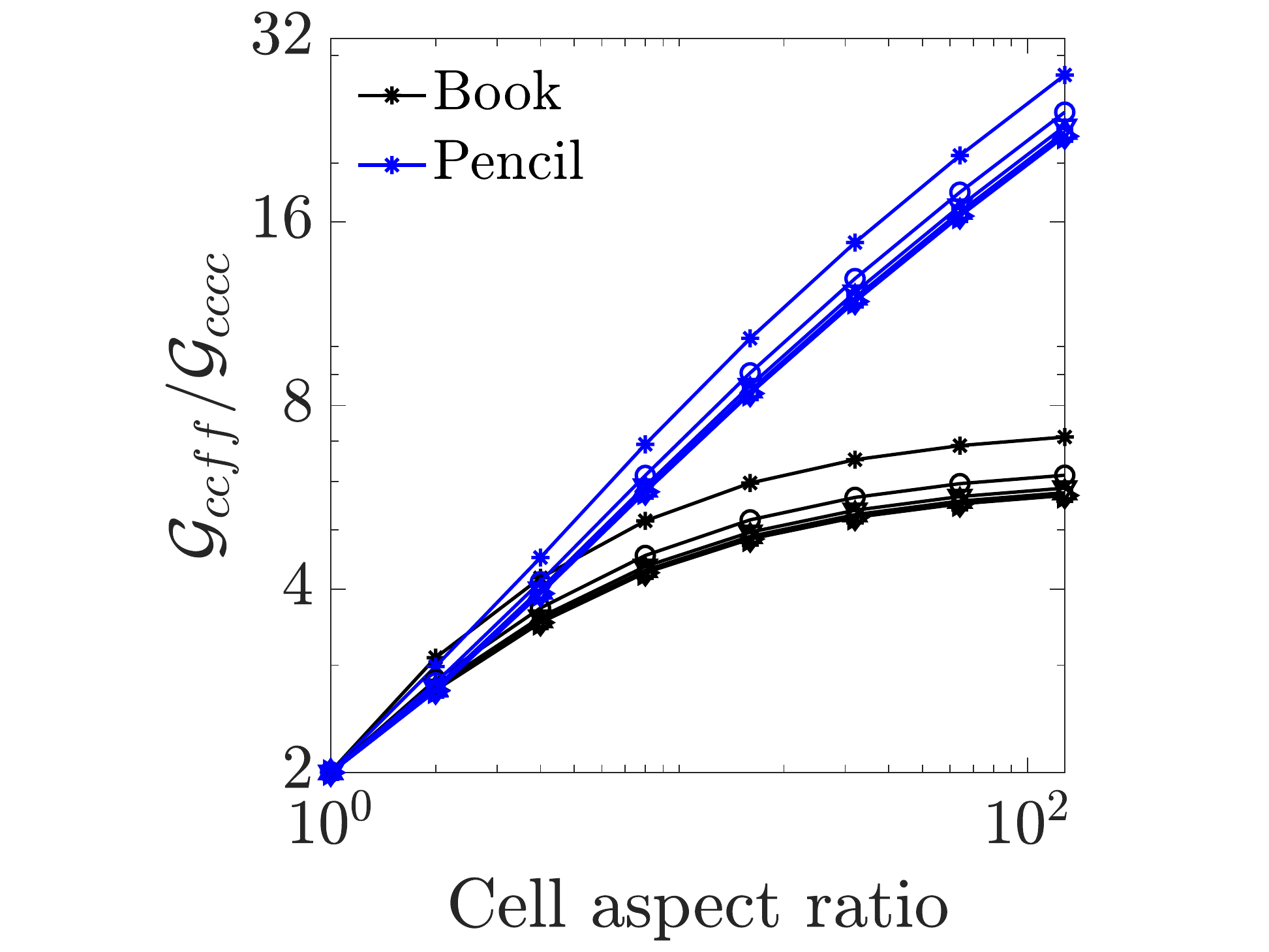}\label{Gc}}\\
 \end{center}
 \caption{Resolution-induced anisotropy in the second moment of the
 resolved velocity gradient tensor $\cG_{ijkl}$ determined from
 (\ref{Gijkl}) based on wavenumber domain
 $\cD^e$ given by (\ref{DefinecDe}). The value of $k_{cc}/k_m$ is 2
   *, 4 $\circ$, 8 $\triangledown$, 16 $\vartriangle$, 32 $\square$, 64
   $\lozenge$, and 128 $\vartriangleright$.
   Subscript ``c'' indicates the coarse direction while ``f'' indicates the fine direction.}
 \label{g_aniso}
 \end{figure}
 

In the case of book and pencil cell resolution, where the resolution
in two directions is the same, the elements of $A$ include four
distinct values. So, the impact of anisotropic resolution on
$\cG_{ijkl}$ can be fully described by three ratios of its non-zero
elements, as shown in Fig.~\ref{g_aniso} as a function of the
resolution aspect ratio.  Plotted are the ratios
$\cG_{ffff}/\cG_{cccc}$, $\cG_{ffcc}/\cG_{cccc}$ and
$\cG_{ccff}/\cG_{cccc}$, where subscripts $f$ and $c$ indicate the fine
and course resolution directions, respectively. If $\cD$ is isotropic
(spherically symmetric), these ratios have values 1, 2 and 2
respectively. Notice that for book cells the anisotropy of $\cG$ saturates with
increasing cell aspect ratio, as with the Reynolds stress, but
for pencil cells, it grows like the cell aspect ratio to a power
between 0.4 and 0.6.

\begin{figure}[tp]
 \begin{center}
 
  \subfigure[Book Fine]{\includegraphics
     [width=0.45\linewidth,bb=35 0 486 419,clip]{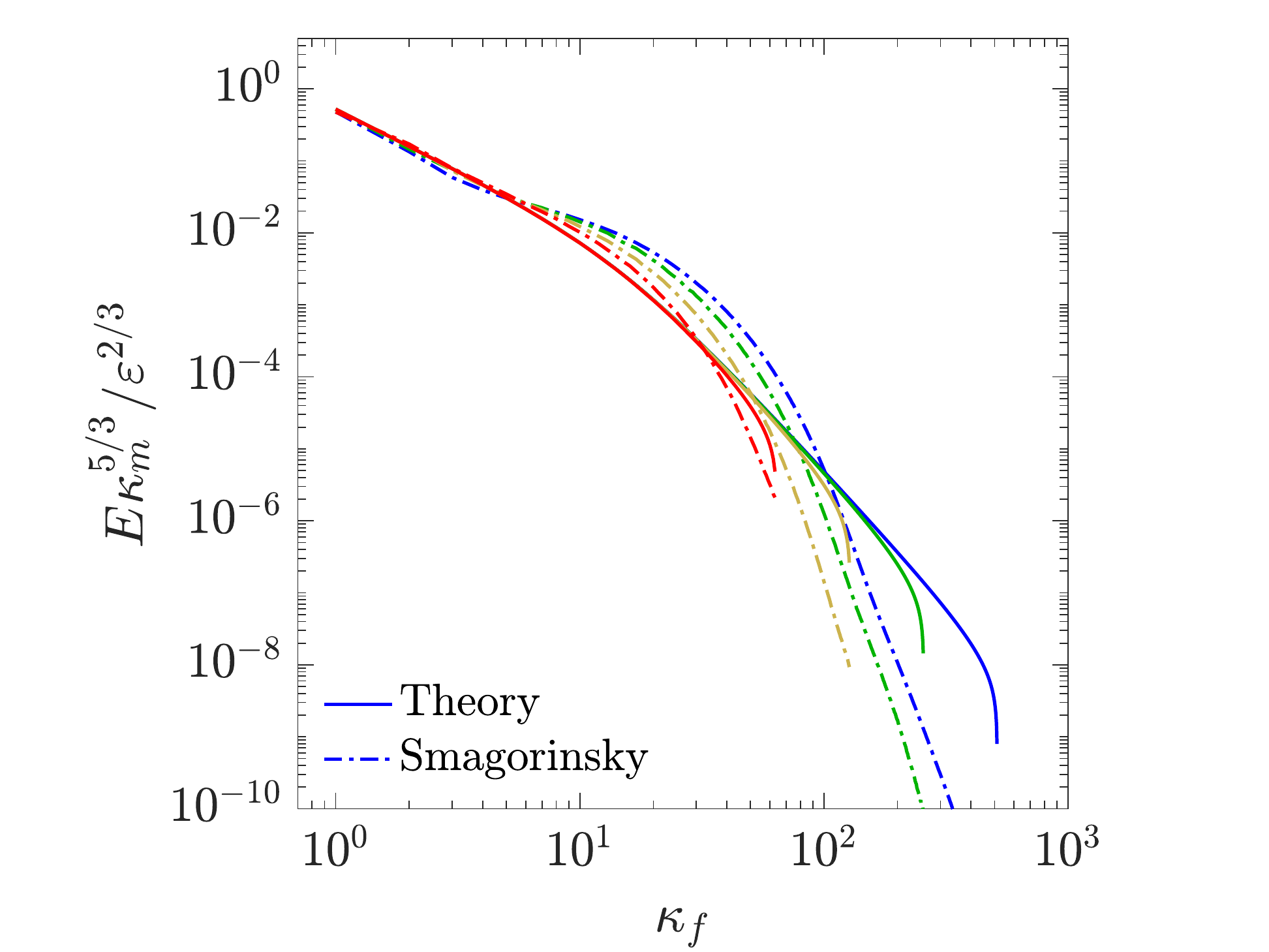} \label{smag_fbook}}
  \subfigure[Book Coarse]{\includegraphics
     [width=0.45\linewidth,bb=35 0 486 419,clip]{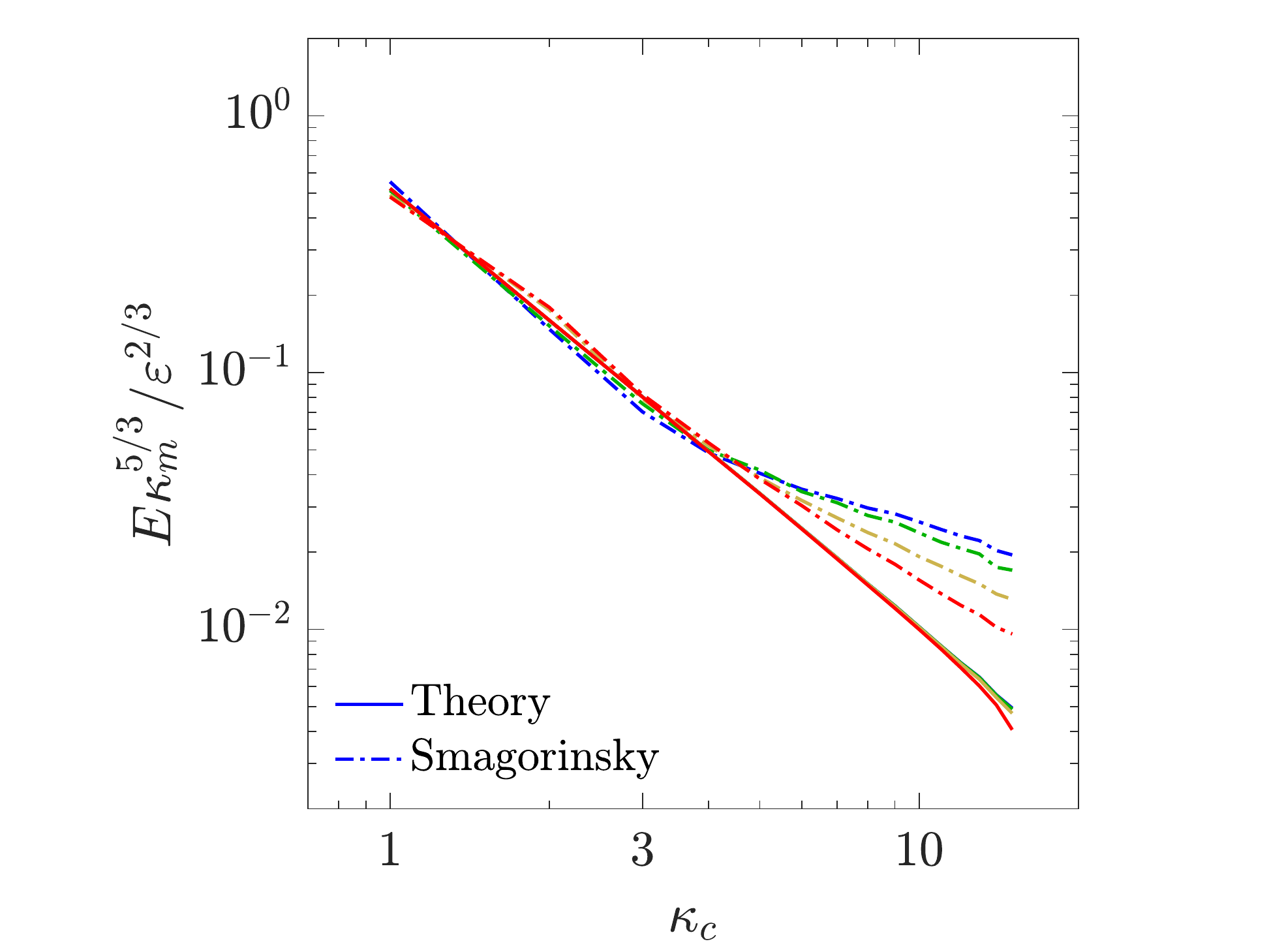} \label{smag_cbook}}\\
Cell aspect ratio: {\color{red} 4}, {\color{dy} 8},
 {\color{dg} 16}, {\color{blue} 32}\\
\subfigure[Pencil Fine]{\includegraphics
     [width=0.45\linewidth,bb=35 0 486 419,clip]{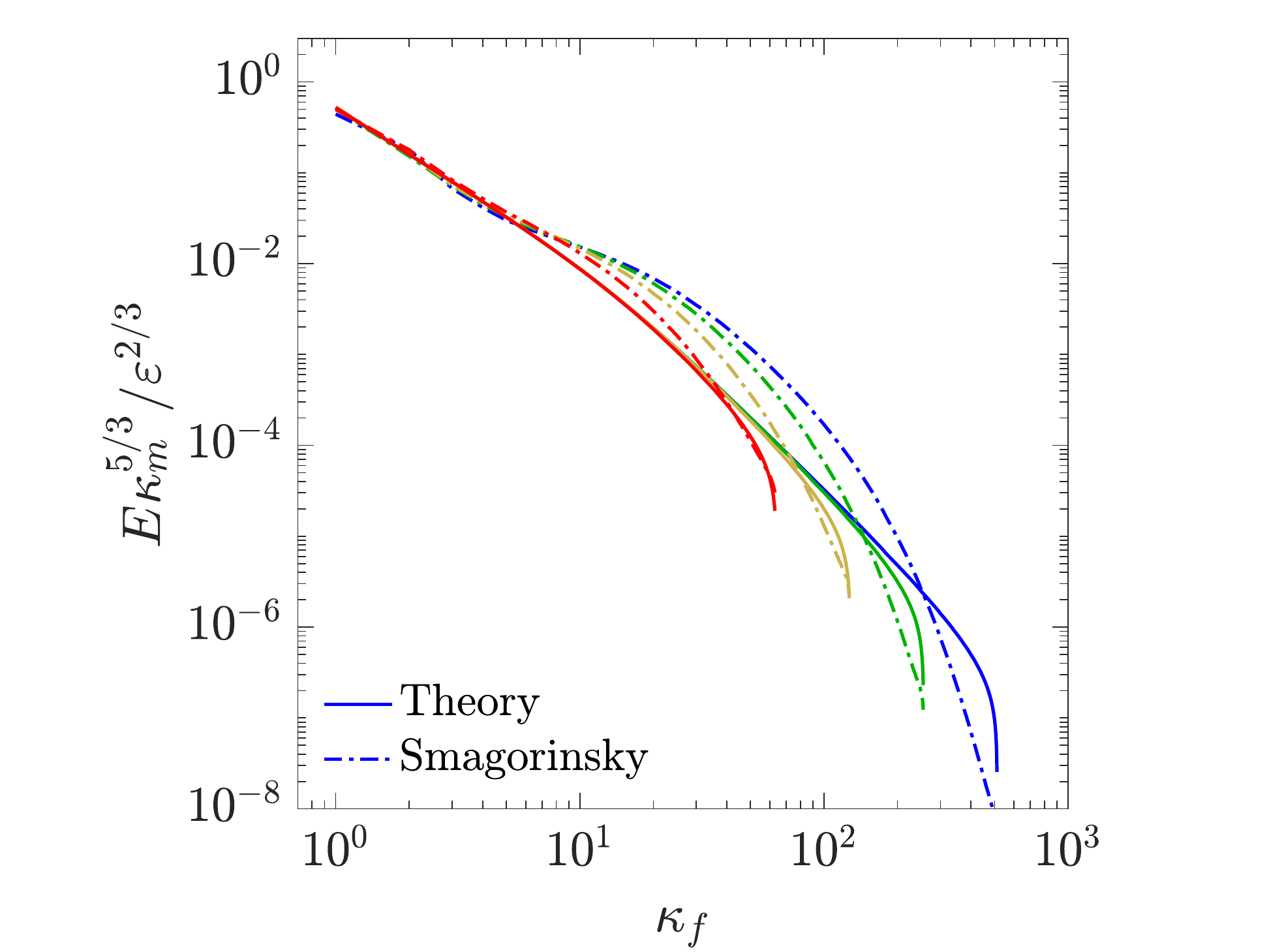} \label{smag_fpencil}}
  \subfigure[Pencil Coarse]{\includegraphics
     [width=0.45\linewidth,bb=35 0 486 419,clip]{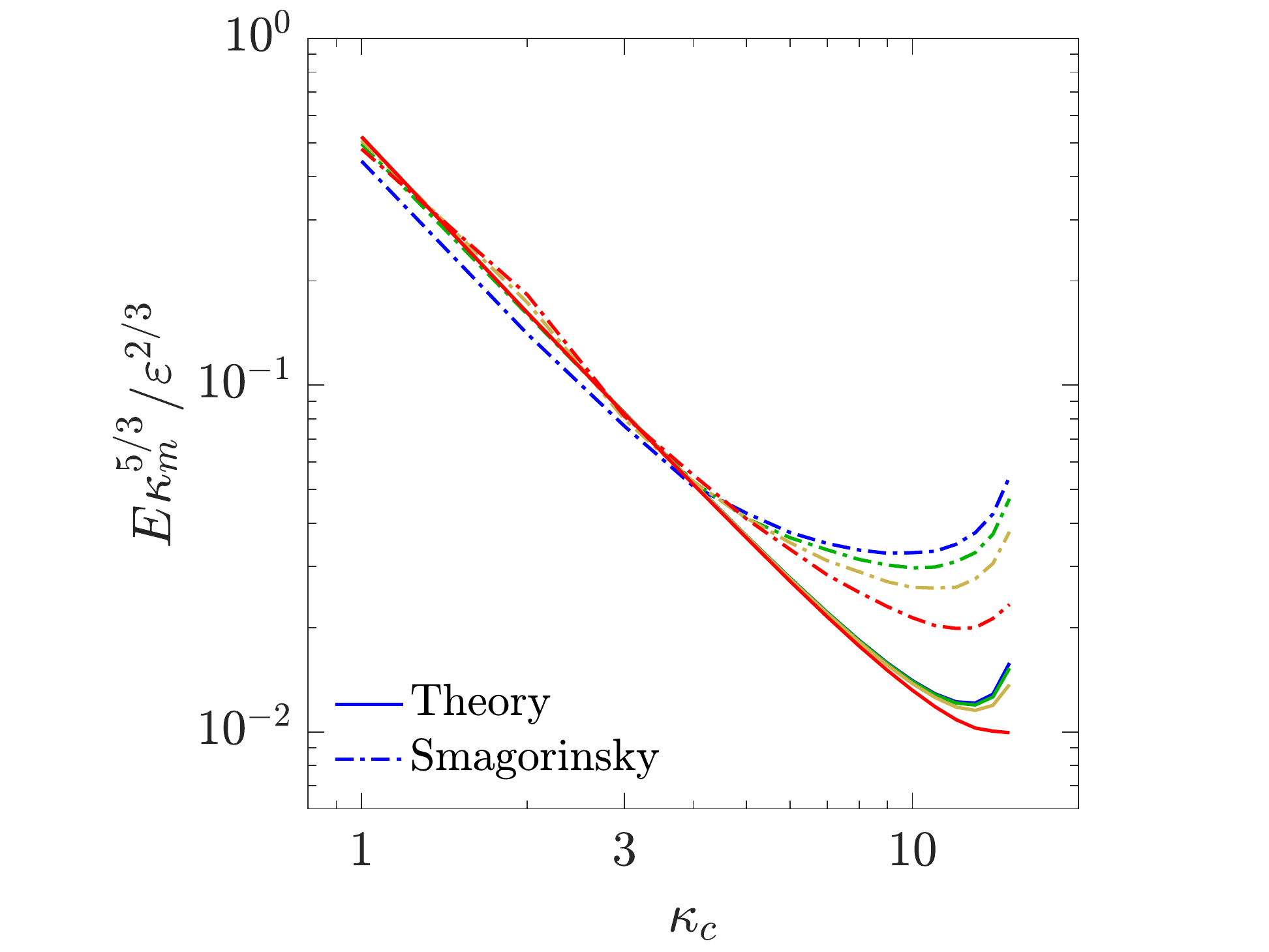} \label{smag_cpencil}} \\
 \end{center}
\caption{One-dimensional energy spectra $E$, from LES with the standard
 Smagorinsky model and anisotropic resolution, compared with the
 equivalently filtered $|\bkappa|^{-5/3}$ Kolmogorov inertial range
 spectra.}  \label{smag}
\end{figure}

 
The consequences of neglecting resolution anisotropy and the resulting
gradient anisotropy can be demonstrated in LES using the basic
Smagorinsky model \cite{smag:1963} applied to simulations with
anisotropic resolutions.  In these simulations, and elsewhere in this
paper, LES of forced homogeneous isotropic turbulence at infinite $Re$
is performed with a modified version of the dealiased pseudo-spectral
code \emph{Poongback}\cite{LeeMoser2015} using cell aspect ratios
ranging from 4 to 32 in a $2\pi$ box. Negative viscosity forcing is performed over a
band of wavenumbers with magnitudes $|\bkappa|=(0.0,2.0]$ with $P_{in}=0.103$. The length scale in the Smagorinsky model was
taken from the cell volume, and the Smagorinsky constant was set to $C_s=0.013$ ($\nu_t = C_s\sqrt{2\bar{S}_{ij}\bar{S}_{ij}}\delta^2_{vol}$)
by optimizing results with isotropic resolution.
One-dimensional spectra in the fine and coarse
directions from these LES are compared to the same obtained from an
equivalently filtered $|\bkappa|^{-5/3}$ Kolmogorov inertial range
(Fig. \ref{smag}). Spectra are averaged over
10 fields spanning at least four eddy turnover times. For both pencil
and book cells, the LES fine-direction spectra have excess energy in
the mid wavenumbers centered around the coarse cutoff wavenumber along with a rapid rolloff at high wavenumbers. This high wavenumber energy deficit is most pronounced for book cells.  In the coarse-direction spectra, the LES
exhibit excess energy at the cutoff, up to a factor of approximately
six.  This ``energy pile-up'' at the coarse cutoff appears to be
saturating at the highest aspect ratio of 32.

Note that using the dynamic approach \cite{germ:1991} would not improve these results as 
the effects of resolution anisotropy would be lost to averaging over all homogeneous 
directions.  Indeed, it has been shown \cite{scot:1996b} that test filtering should be performed 
isotropically corresponding to the coarsest grid spacing to obtain decent energy spectra up to the minimum cutoff wavenumber with 
a large roll-off thereafter when using 
the dynamic approach and anisotropic grids.  It may be possible to construct a tensor-based dynamic coefficient to 
correct this issue but such an approach has not been explored here.  The anisotropy 
correction of Scotti \emph{et al.} \cite{scott:1993} is considered later.

Since the turbulence is homogeneous, this test is insensitive to any
errors in representing the subgrid contribution to the Reynolds
stress. Therefore, it seems likely that the errors in the energy
spectra are purely a result of incorrectly modeling the dissipation
anisotropy.  This hypothesis is examined later (see Sec.~\ref{sec:eval}).  With
the energy input fixed through low wavenumber stirring, the turbulent
field and its gradients must adjust to come into equilibrium and
produce a total dissipation equal to the energy injection rate. The
deficiencies in the model require that resolved turbulence be
distorted to reach this equilibrium, thus the inconsistent spectra. In
a more realistic flow scenario, this distortion of the turbulence may
have significant consequences. For example, amplifying the energy
content near the course grid scale could result in anomalously large 
turbulent transport, as such lower wavenumbers possess more turbulent kinetic energy.  The result would be incorrect
prediction of mean quantities. Therefore, here we aspire to pose
models that do not distort the resolved turbulence spectra, even in
the presence of resolution anisotropy.

\section{A simple anisotropic subgrid model}
\label{sec:M43}
To treat resolution anisotropy in subgrid modeling, we begin by
relaxing the assumption of an isotropic eddy viscosity, so that the
diffusivity can have directional dependence. With an anisotropic eddy
viscosity, the standard subgrid model formulation is modified, with
the simplest model for the deviatoric portion of the subgrid stress
tensor $\tau_{ij}$ in terms of a symmetric second rank tensor eddy
diffusivity $\nu_{ij}$ and the velocity gradient tensor given by
\begin{equation}
-\big{(}\tau_{ij}-\tfrac{2}{3}k_{sgs}\delta_{ij}\big{)}\approx\nu_{jk}\partial{}_k\overbar{u}_i+\nu_{ik}\partial{}_k\overbar{u}_j - \tfrac{2}{3}\nu_{kl}\partial{}_l\overbar{u}_k\delta_{ij},
\label{rs}
\end{equation}
where $k_{sgs}=\tfrac{1}{2}\tau_{kk}$.
Note that the anisotropy of $\nu_{ij}$ acts on the derivative
components of the gradient tensor, not the velocity components. This is
consistent with the fact that the resolution anisotropy introduces
anisotropy in scale. Obviously, if $\nu_{ij}$ is isotropic, this form
reduces to a standard Boussinesq eddy viscosity model. When $\nu_{ij}$
is anisotropic, the model for $\tau_{ij}$ depends on both the strain
rate tensor and the rotation rate tensor, unlike a Boussinesq model.

Second, to support the development of an anisotropic eddy viscosity,
we introduce a resolution tensor to express the anisotropy of the
resolution. The resolution tensor, $\mathcal{M}_{ij}$, is
formally the symmetric part of the Jacobian defining the mapping of a
unit cube to a resolution cell, or equivalently, the square-root of
the cell metric tensor \cite{syng:1949}.  The eigenvalues
$\lambda_i^\cM$ of
$\mathcal{M}_{ij}$ therefore represent the size of a resolution cell
in the principal directions while its
eigenvectors define those principal directions. 
Common grid measures are invariants or eigenvalues of
$\mathcal{M}_{ij}$; for example,
$\delta_{min}=\min_i\lambda_i^\mathcal{M}$,
$\delta_{diag}=(\mathcal{M}_{ij}\mathcal{M}_{ji})^{1/2}$,
$\delta_{vol}=(\mathbf{\det}(\mathcal{M}))^{1/3}$, which are the
minimum dimension, the diagonal and the cube root of the volume of a
resolution cell, respectively.  The resolution tensor is a more
complete representation of the resolution than scalar measures, so
incorporating it into a subgrid model allows the model to retain
information about the anisotropy of the resolution. Using a tensor
representation of resolution ensures that models constructed from
it will be consistent, independent of the orientation of local
coordinate systems. 


Of primary importance in LES is the LES ``dissipation,'' that is the
transfer of energy from the resolved to unresolved scales. Because the
LES resolution is anisotropic, variations of resolved velocity in
different directions contribute differently to the dissipation. This
scale anisotropy of the dissipation can be characterized by the total dissipation 
tensor $\hat{\varepsilon}_{ij}$ defined as 
\begin{equation}
  \hat{\varepsilon}_{ij}=-\tfrac{1}{2}\big(\la\partial_j\bar{u}_k\tau_{ik}\ra+\la\partial_i\bar{u}_k\tau_{jk}\ra\big).
  \label{eq:epsilonij}  
\end{equation}
To capture the anisotropic character of this energy
transfer, the anisotropic eddy viscosity in the model (\ref{rs})
should satisfy
\begin{equation}
\varepsilon_{ij} = \tfrac{1}{2}\big(\la{}\nu_{ik}\partial_j\bar{u}_l\partial_k\bar{u}_l\ra + \la{}\nu_{jk}\partial_i\bar{u}_l\partial_k\bar{u}_l\ra + \la{}\nu_{lk}(\partial_j\bar{u}_l\partial_k\bar{u}_i+\partial_i\bar{u}_l\partial_k\bar{u}_j)\ra\big) -\tfrac{2}{3}\la\nu_{kl}\partial_k\bar{u}_l\bar{S}_{ij}\ra,
\label{e1}
\end{equation} 
where $\varepsilon_{ij} = \hat{\varepsilon}_{ij}
- \tfrac{2}{3}\la{}k_{sgs}\bar{S}_{ij}\ra$.  This modified dissipation
tensor characterizes the contribution of the deviatoric part of the
subgrid stress $\tau$, which is the part that the model (\ref{rs})
represents. Further, because of continuity, the isotropic part of
$\tau$ does not contribute to the dissipation, so that the dissipation
$\varepsilon=\hat\varepsilon_{ii}=\varepsilon_{ii}$.
The simplest possible anisotropic eddy viscosity model is one in which
the eddy viscosity does not fluctuate, so that in (\ref{e1}) it can be
moved out of the expected value, resulting in
\begin{equation}
  {\varepsilon}_{ij} = \tfrac{1}{2}\big(\nu_{ik}\cG_{lljk}+\nu_{jk}\cG_{llik} + \nu_{lk}(\cG_{lijk}+\cG_{ljik})\big) - \tfrac{1}{3}\nu_{lk}(\cG_{jkil}+\cG_{ikjl}).
\label{eq:epsG}
\end{equation}
Furthermore, in isotropic turbulence, all anisotropy arises due to the
anisotropy of resolution. Therefore, the tensors $\varepsilon_{ij}$,
$\nu_{ij}$ and $\cM_{ij}$ must all have the same
eigenvectors and for the Cartesian tensor product filtering used here,
these are just the coordinate basis. In this basis, $\varepsilon_{ij}$
and $\nu_{ij}$ are diagonal with the eigenvalues
$\lambda^\varepsilon_\alpha=\varepsilon_{\alpha\alpha}$ and
$\lambda^\nu_\alpha=\nu_{\alpha\alpha}$
on the diagonals, and $\cG_{ijkl}$ has the characteristics described
in Sec.~\ref{sec:aniso}. This allows the $\alpha$ eigenvalue
$\lambda_\alpha^\nu$ of $\nu_{ij}$
to be determined in terms of the eigenvalues
$\lambda_\alpha^\varepsilon$  of $\varepsilon_{ij}$ by solving the following coupled system
of three linear equations 
\begin{equation}
\lambda^{{\varepsilon}}_{\alpha}=\lambda^\nu_\alpha\cG_{jj\alpha\alpha}+\sum_\beta\lambda^\nu_\beta\cG_{\beta\alpha\alpha\beta} - \tfrac{2}{3}\sum_\beta\lambda^\nu_\beta\cG_{\alpha\beta\alpha\beta},
\label{e2}
\end{equation} 
where again, no summation is implied for repeated Greek indices.  The
resulting anisotropic eddy viscosity is \emph{ a priori} consistent
with the scale anisotropy of the dissipation ${\varepsilon}_{ij}$.
Unfortunately, while we have an evaluation of $\mathcal{G}_{ijkl}$
from inertial range theory, we have no such simple evaluation for
${\varepsilon}_{ij}$.  A model for the three-point third-order correlation has been 
formulated \cite{chang:2007}, and in principle it could be used to
develop a representation for $\varepsilon_{ij}$ for a given definition
of the resolved turbulence, but that is out of
scope of the current paper. Instead, we consider the form of the model
relevant for isotropic resolution, and use its form as a guide to
generalizing to anisotropic resolution.  





When $\cD$ is spherically symmetric including wavenumbers with
magnitudes ranging from $\kappa_m$ to $\kappa_c$, the integral in (\ref{Gijkl})
can be performed over spheres to obtain
\begin{equation}
\mathcal{G}_{\alpha\alpha\alpha\alpha}=\tfrac{4}{5}C_k\varepsilon^{2/3}\big{(}\kappa_c^{4/3}-\kappa_{m}^{4/3}\big{)}
\label{Giso}
\end{equation}
which, due to isotropy and the constraints described in Sec.~\ref{sec:aniso},
completely determines the $\kappa_c$ dependence of $\cG_{ijkl}$. All other components are
proportional to $\cG_{\alpha\alpha\alpha\alpha}$, with the constant of
proportionality depending on the component. Therefore the solution of
(\ref{e2}) in this case must be
\begin{equation}
  \lambda^\nu=C\varepsilon^{1/3}(\kappa_c^{4/3}-\kappa_m^{4/3})^{-1},
  \label{nuiso}
\end{equation}
where $C$ is a constant, which in principle is determined in terms of $C_k$, and $\nu_{ij}=\lambda^\nu\delta_{ij}$ is
isotropic.
Since $\kappa_c=\pi/\Delta$ with $\Delta=\lambda^\cM$
and $\kappa_m=2\pi/L$ with $L$ either the domain size or
proportional to the integral scale, a consistent generalization to
anisotropic resolution is
\begin{equation}
\lambda^\nu_\alpha=C(\hat{\cM})\varepsilon^{1/3}((\lambda^\mathcal{M}_\alpha)^{-4/3}-(L/2)^{-4/3})^{-1},
\label{M43_basic}
\end{equation}
where now $C(\cM)$ generally depends only on the
invariants of the
scaled resolution tensor $\hat{\cM}=\cM/\delta$, where
$\delta=\min_i \lambda^\cM_i$; that is, only on the anisotropy of
$\cM$.
If the $\lambda^\cM$ are
sufficiently small compared to $L$, then the $L$ term can be neglected
yielding the basic M43 model
\begin{equation}
\nu_{ij}=C(\hat{\cM})\varepsilon^{1/3}\mathcal{M}^{4/3}_{ij}.
\label{M43}
\end{equation}
When the $\lambda^\cM$ are not small enough compared to $L$, the more
general ``low-k'' version, consistent with (\ref{M43_basic}) is
obtained by replacing $\cM$ in (\ref{M43}) with a modified resolution tensor $\cM^*$
given by 
\begin{equation}
\cM_{ij}^*=(\cM^{-4/3}-(L/2)^{-4/3}\cI)^{-3/4}_{ij}
\end{equation}
The coefficient $C(\cM)$ is determined by requiring that
when applied to the filtered theoretical spectrum the model dissipates
energy at the rate $\varepsilon$ (see Appendix A). Thus, the coefficient
is determined theoretically in terms of the Kolmogorov constant $C_k$,
consistent with the filter and numerical representation used in the LES.
Also provided in Appendix A is a fit used here to represent the functional
dependence of $C$ on the eigenvalues of $\cM$. Consistent with the
Fourier spectral representation used in the LES, $C(\cM)$ is
determined here based on $\cG$ computed on the $\cD^c$ domain defined
in (\ref{DefinecDc}).

\section{Model tests}
\label{sec:eval}

In this section, the proposed M43 model, its low-k variant and the AMD
model of \cite{roze:2015} are tested in detail in infinite Reynolds
number LES of isotropic turbulence with anisotropic resolution. 
For the modified Smagorinsky model of Scotti \cite{vrem:2004} and the
Vreman model \cite{scott:1993}, which also include a dependence on the
resolution anisotropy, a high-aspect ratio example is
included to demonstrate that they exhibit similar deficiencies as the
basic Smagorinsky model (Fig. \ref{smag}).

The AMD model and a subtle technical issue related to its
formulation are discussed briefly in Appendix B. Here we note that
as written in \cite{roze:2015} (equation 23), the model is tensorially
inconsistent. However, it can be recast in terms of the resolution
tensor $\cM$ to be tensorially consistent, while being equivalent to the
formulation in \cite{roze:2015} for rectilinear grids with the grid cells
aligned with the coordinate directions (see Appendix B). It is this
recast form of AMD that we evaluate here. Because the numerical
representation used here is different from that in \cite{roze:2015},
the AMD constant was optimized in an LES with isotropic resolution (see
Sec.~\ref{sec:nofluc}) to
obtain a value of 0.236 rather than the suggested value of 0.212 in \cite{roze:2015}.

Note that the AMD model and the M43 model described in Section \ref{sec:M43} are
fundamentally different. The M43 model poses a tensor eddy viscosity
(\ref{M43_basic})  with the only flow-dependence being the
expected value of the kinetic energy dissipation, $\varepsilon$ . In homogeneous
turbulence, $\varepsilon$ is not spatially varying and in the stationary
flows considered here, it is not time dependent either. Therefore, the
viscosity tensor is a single constant tensor specified entirely \emph{ab initio}. In contrast, the AMD model uses a scalar
eddy viscosity with a non-linear dependence on the local velocity
gradient along with local clipping, and is therefore strongly spatially dependent and discontinuous. With these
stark differences, it is remarkable that these two models produce
energy spectra that are so similar. 
\begin{figure}[tp]
 \begin{center}
 \subfigure[Book Fine]{\includegraphics
    [width=0.45\linewidth,bb=35 0 486 419,clip]{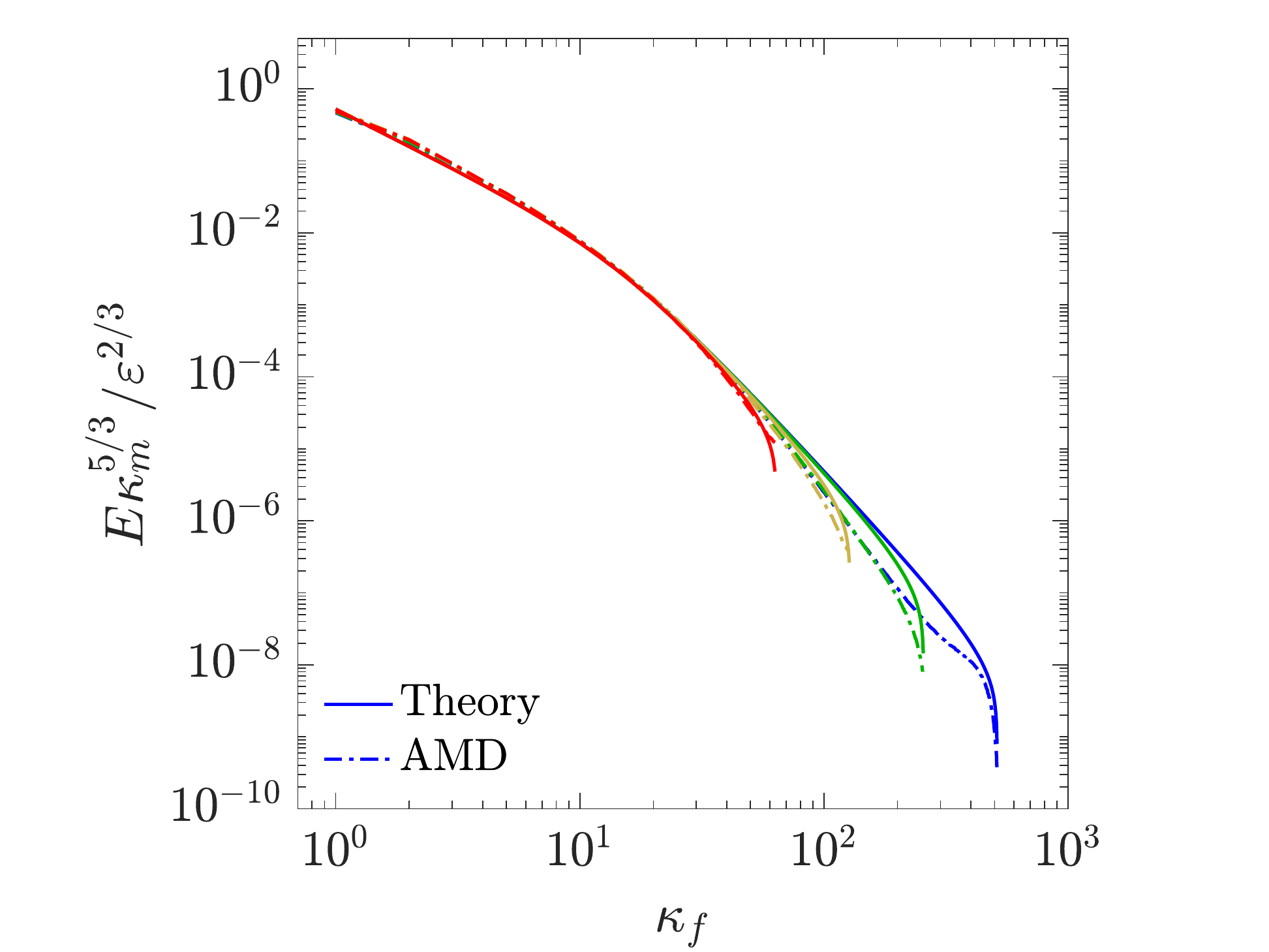}
 \label{amd_fbook}}
 \subfigure[Book Coarse]{\includegraphics
    [width=0.45\linewidth,bb=35 0 486 419,clip]{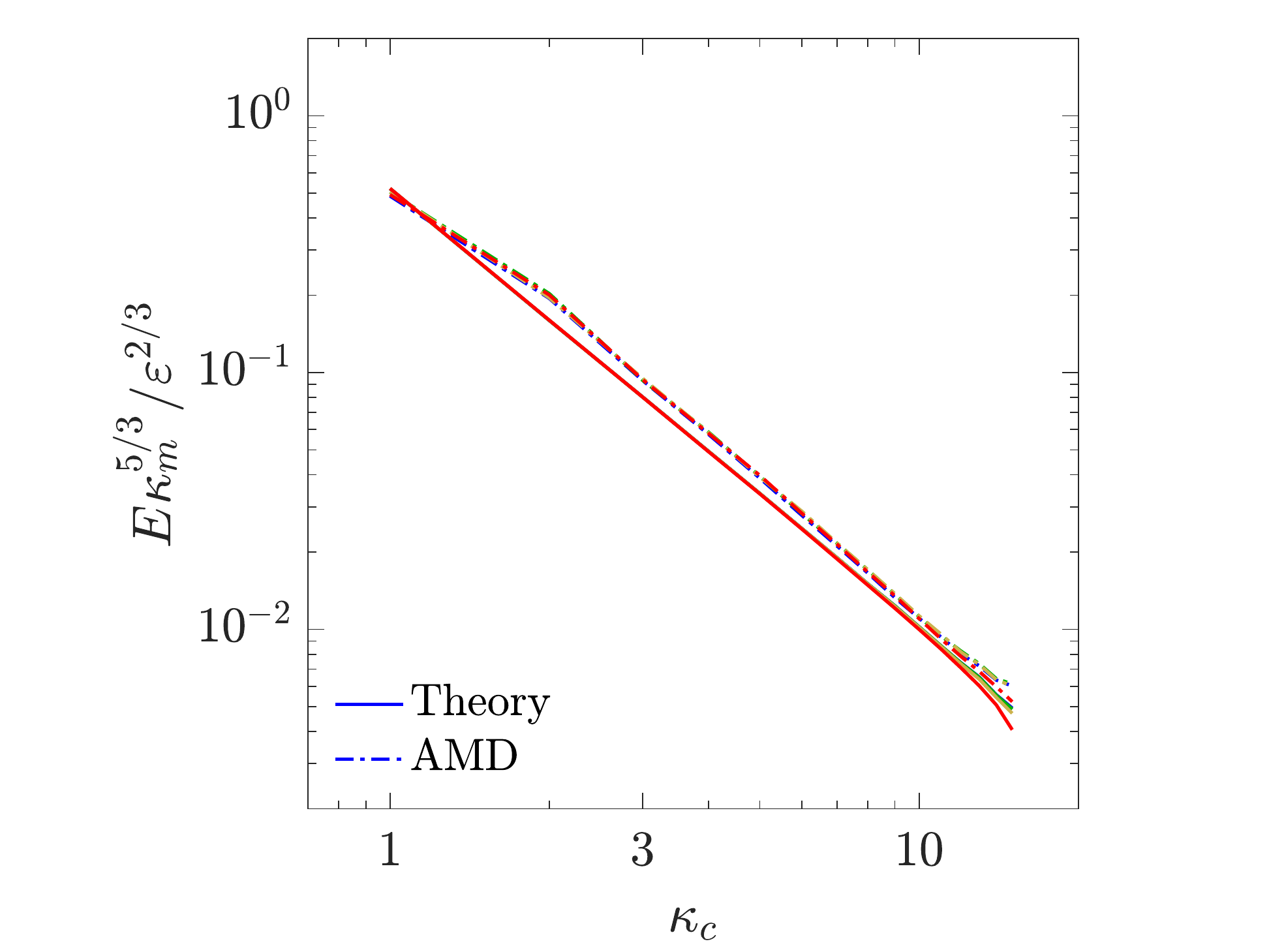}
 \label{amd_cbook}}\\
 Cell aspect ratio: {\color{red} 4}, {\color{dy} 8},
 {\color{dg} 16}, {\color{blue} 32}\\
  \subfigure[Pencil Fine]{\includegraphics
    [width=0.45\linewidth,bb=35 0 486 419,clip]{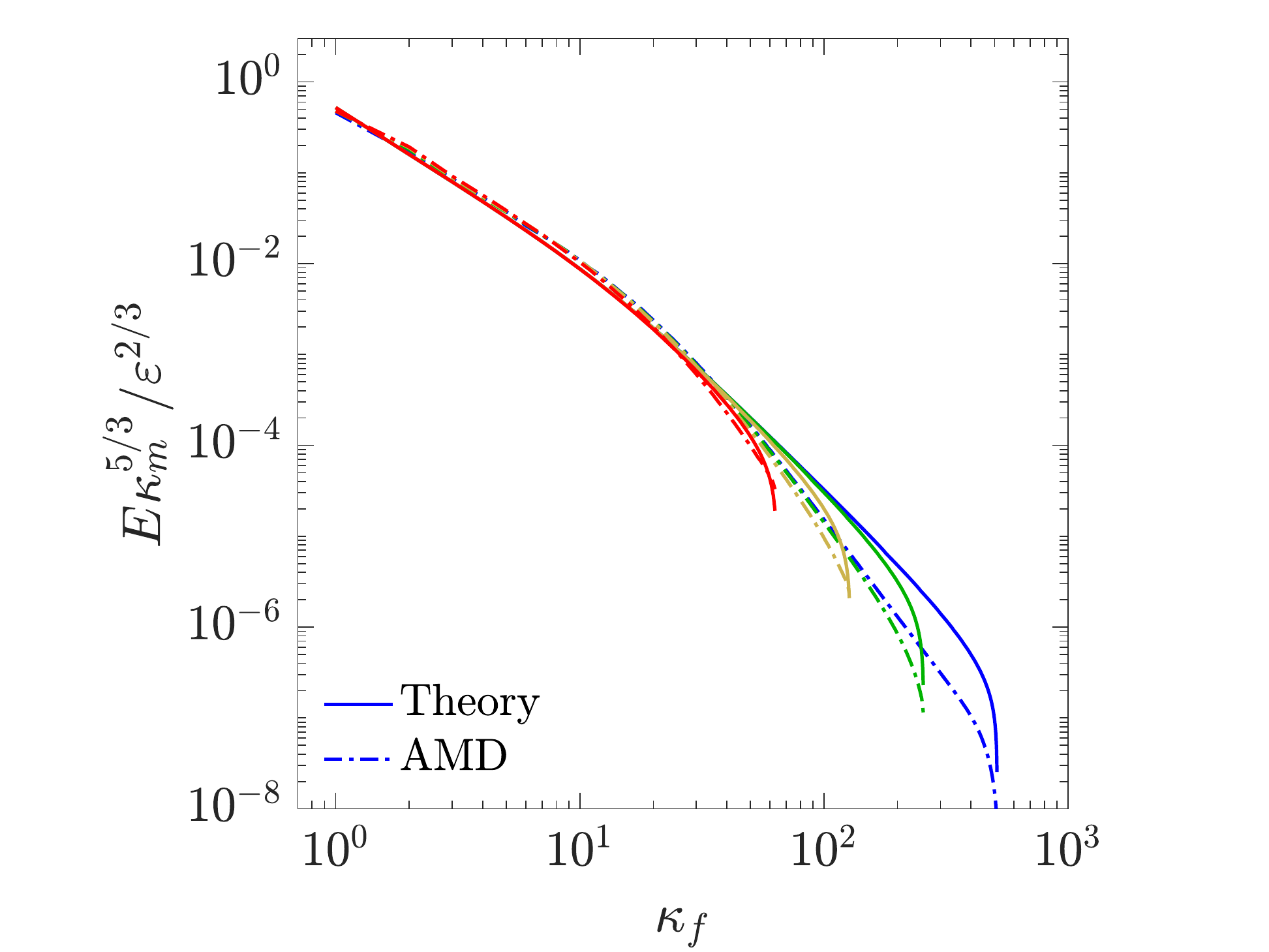}
 \label{amd_fpencil}}
 \subfigure[Pencil Coarse]{\includegraphics
    [width=0.45\linewidth,bb=35 0 486 419,clip]{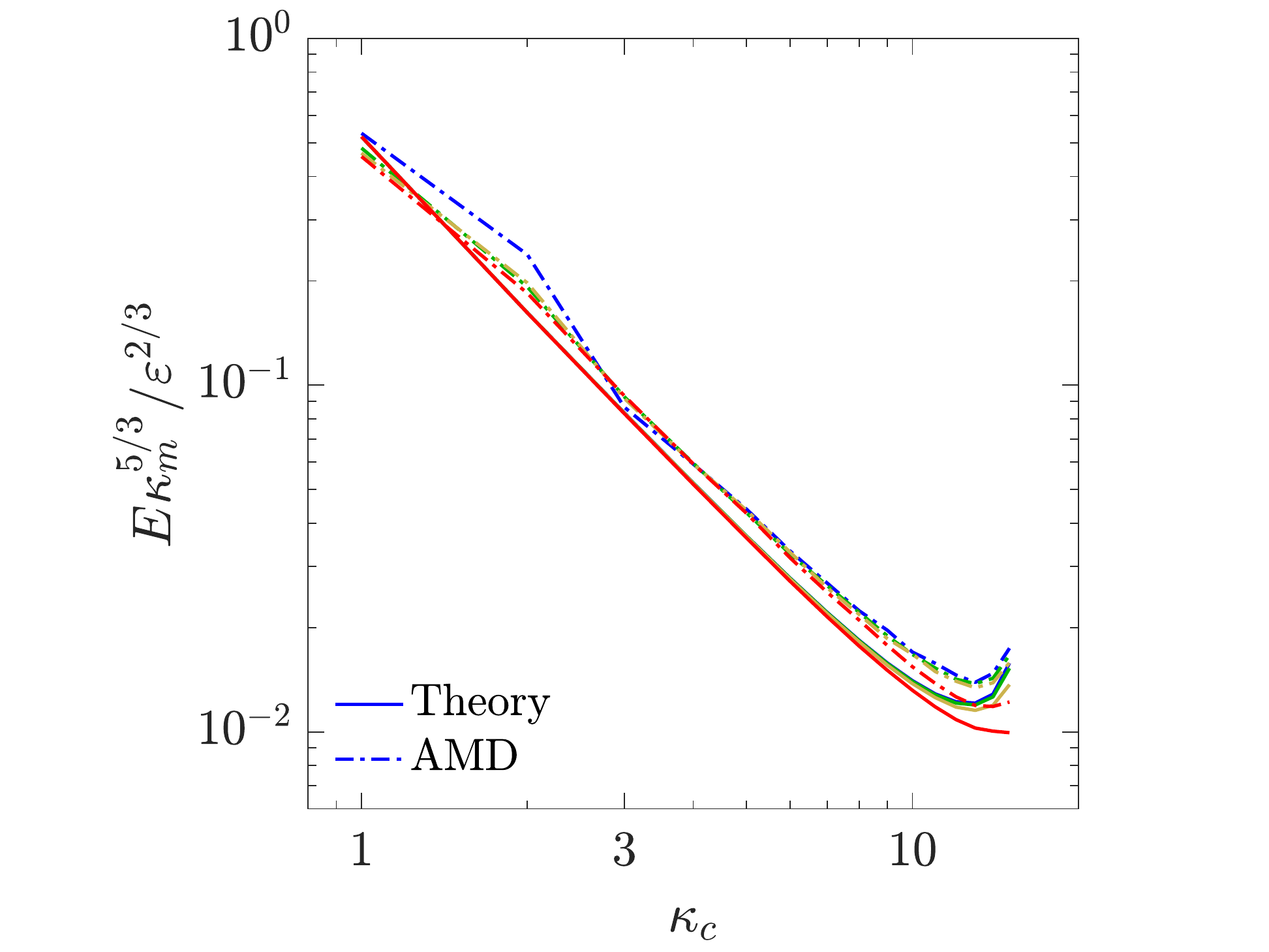}
 \label{amd_cpencil}} \\
\end{center}
\caption{One-dimensional energy spectra $E$ from LES with the AMD
 model and anisotropic resolution, compared with the equivalently
 filtered $|\bkappa|^{-5/3}$ Kolmogorov inertial energy spectra.}
\label{amd}
\end{figure}

\subsection{One-dimensional energy spectra}
\label{sec:eval_spectra}
\begin{figure}[tp]
 \begin{center}
 \subfigure[Smagorinsky]{\includegraphics
    [width=0.45\linewidth,bb=35 0 486 419,clip]{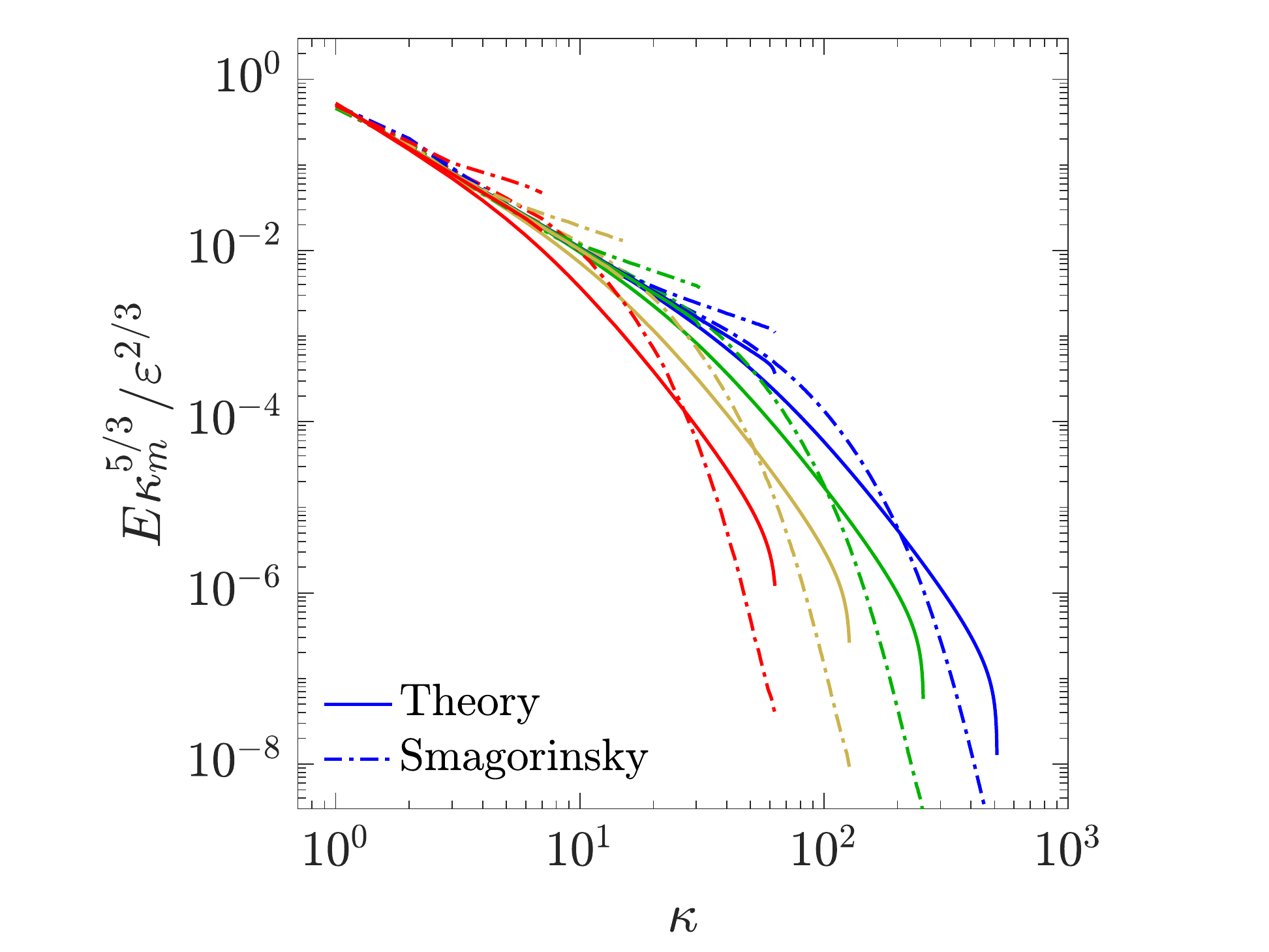}
    \label{smagL}}
 \subfigure[AMD]{\includegraphics
    [width=0.45\linewidth,bb=35 0 486 419,clip]{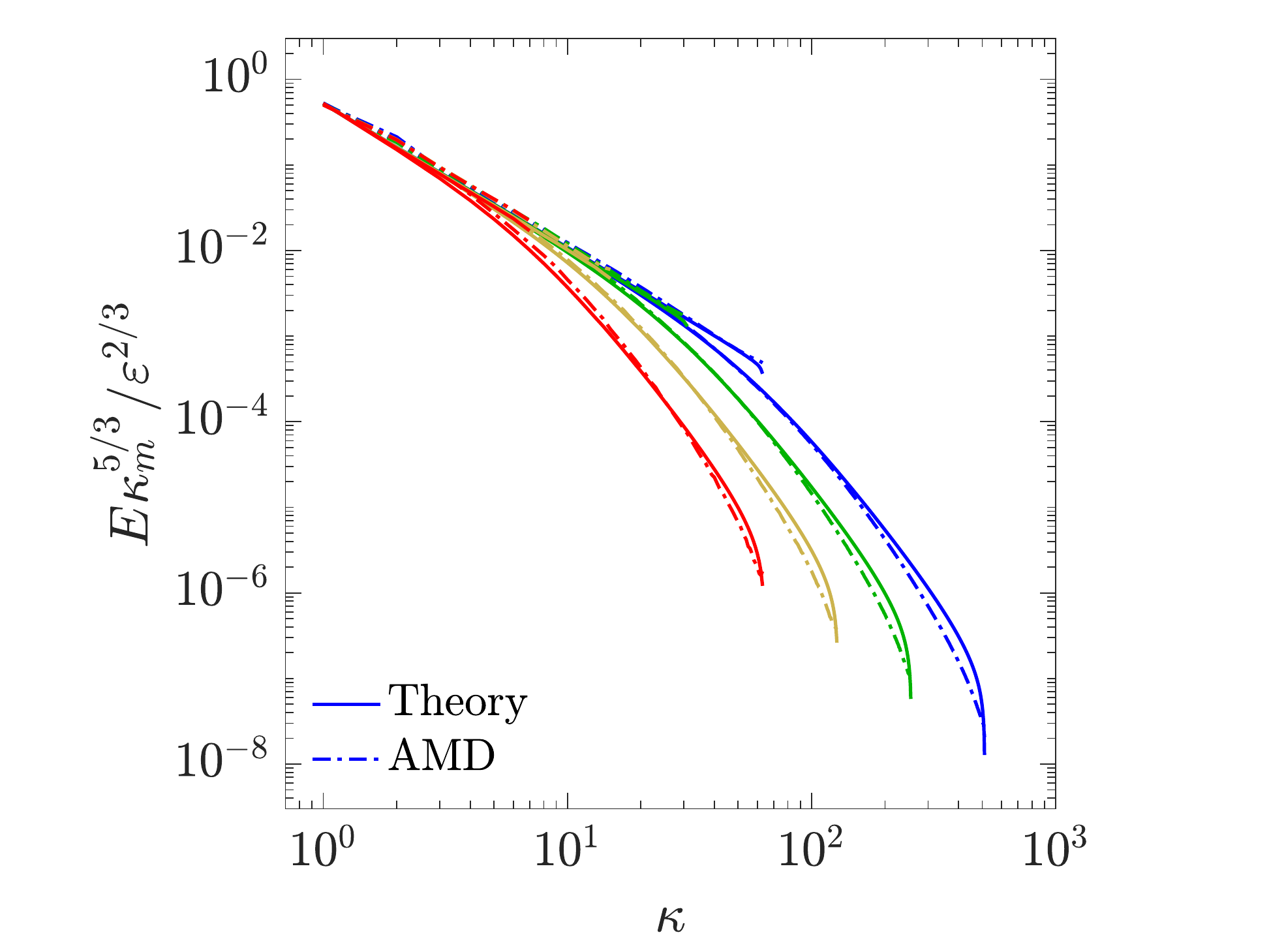}
    \label{amdL}} \\
 Values of $\kappa_{cc}/\kappa_m$ are: {\color{red} 8},
 {\color{dy} 16}, {\color{dg} 32}, {\color{blue} 64}\\
 \subfigure[M43]{\includegraphics
    [width=0.45\linewidth,bb=35 0 486 419,clip]{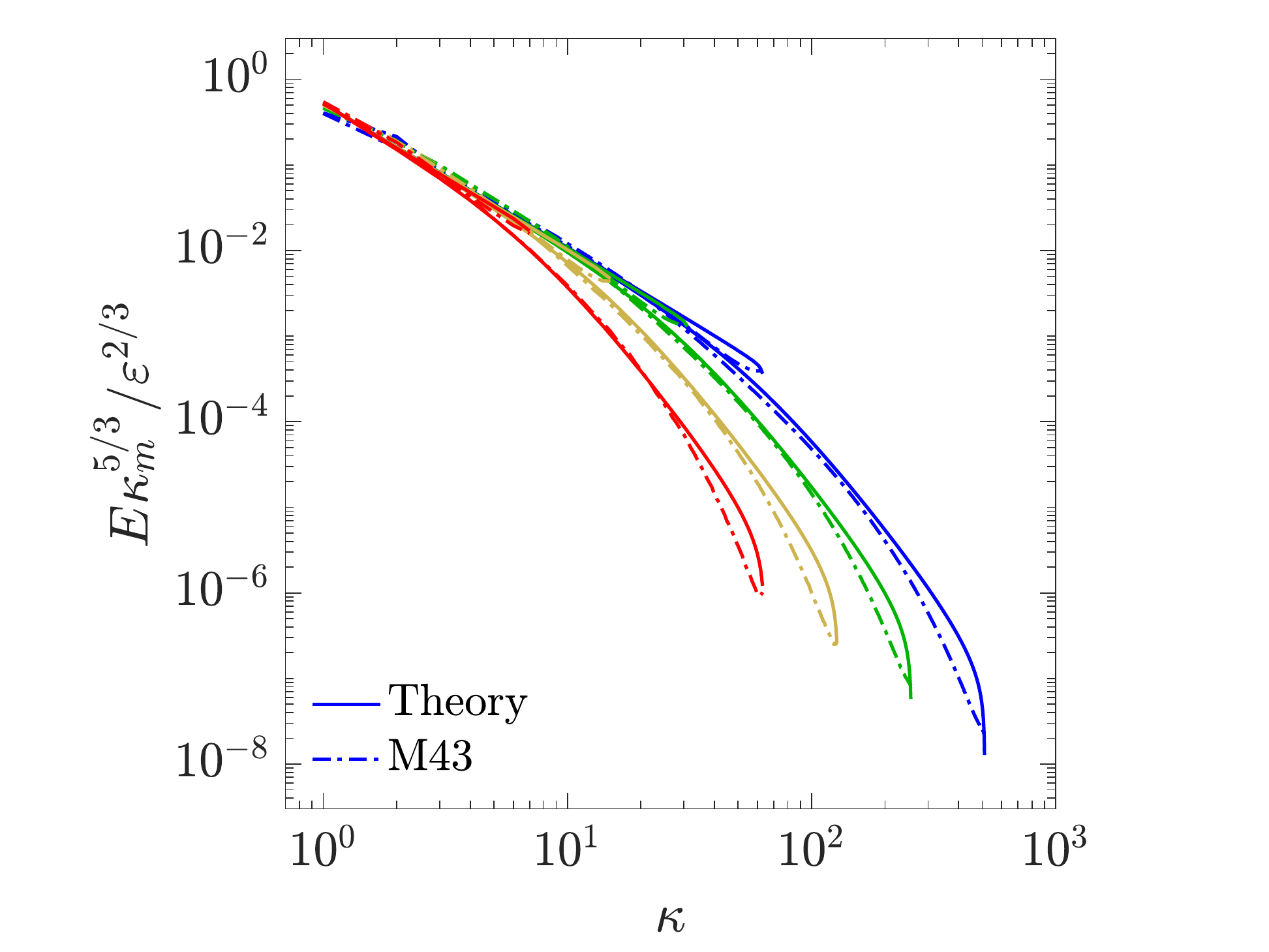}
    \label{m43L}}
 \subfigure[M43 low k]{\includegraphics
    [width=0.45\linewidth,bb=35 0 486 419,clip]{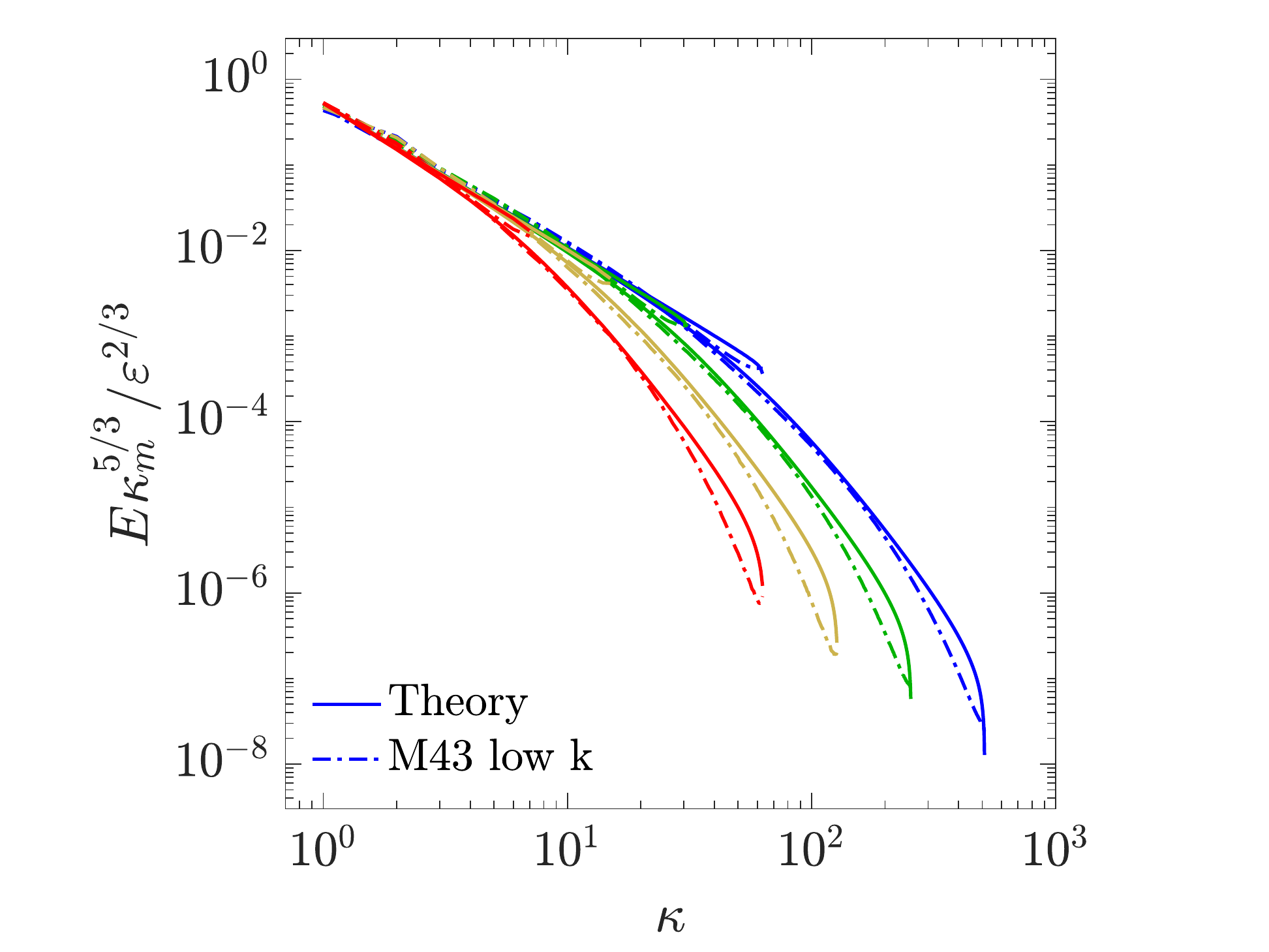}
    \label{m43bLl}} \\
 \end{center}
\caption{One-dimensional energy spectra $E$ from LES with the
 Smagorinsky, AMD, M43 and M43 low-k models and anisotropic resolution
 with aspect ratio 8 and varying values of $\kappa_cc/\kappa_m$,
 compared with the equivalently filtered $|\bkappa|^{-5/3}$ Kolmogorov
 inertial energy spectra. Shown
 are spectra in both the fine and coarse directions.}
 \label{Lratio}
\end{figure}

\begin{figure}[tp]
 \begin{center}
 \subfigure[Book]{\includegraphics
    [width=0.45\linewidth,bb=35 0 486 419,clip]{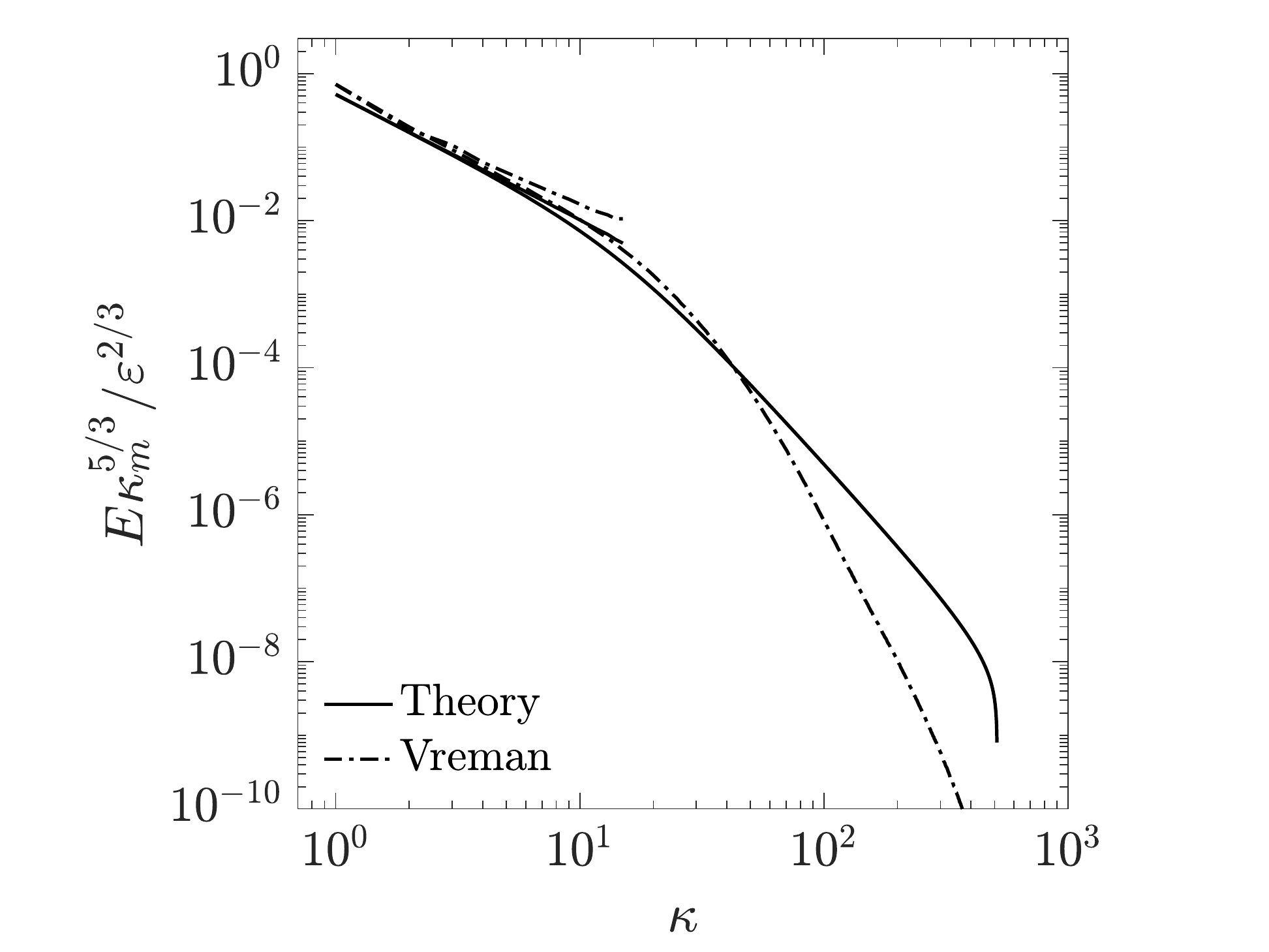}
    \label{vre_b}}
 \subfigure[Pencil]{\includegraphics
    [width=0.45\linewidth,bb=35 0 486 419,clip]{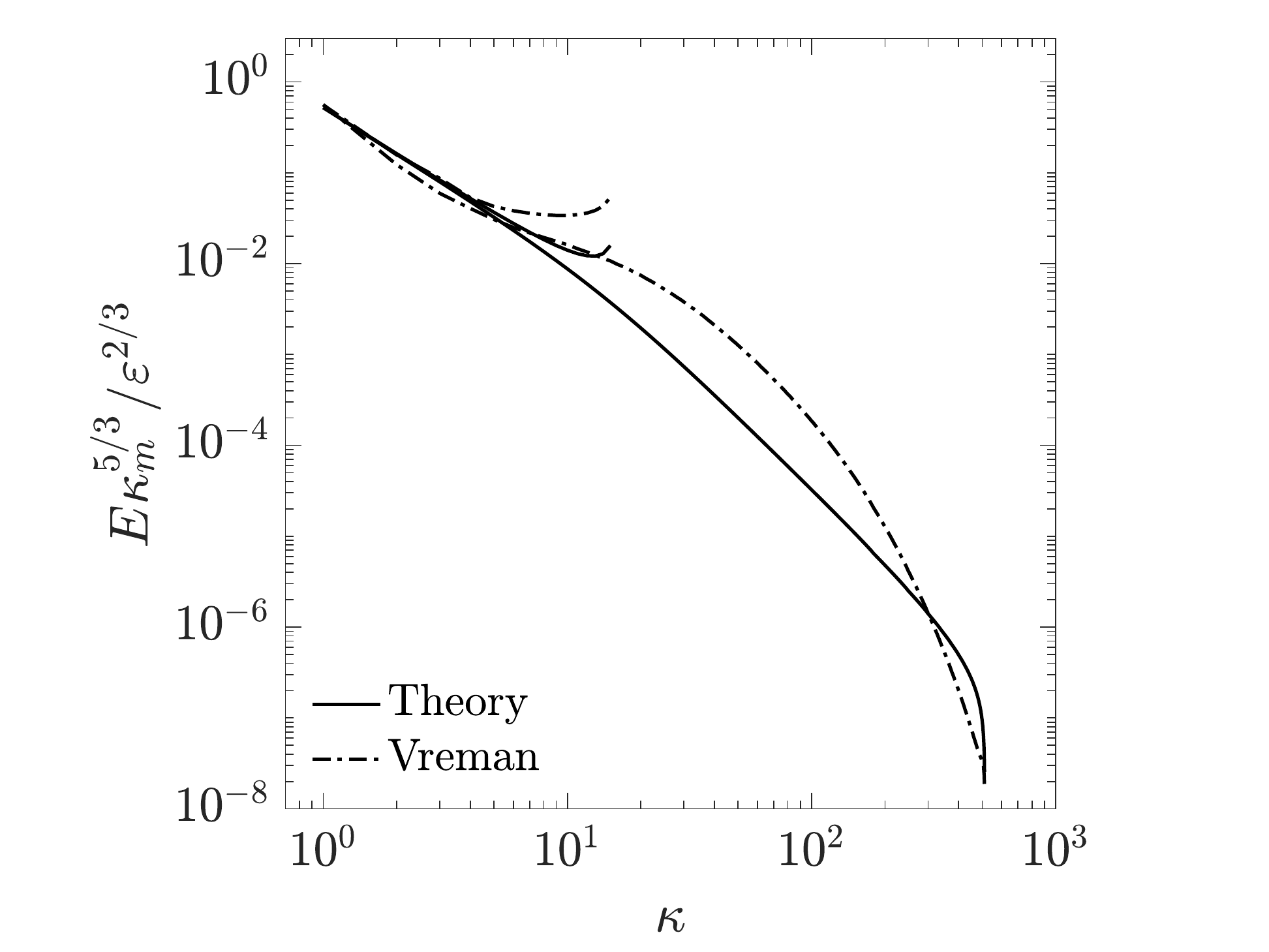}
    \label{vre_p}} \\
 \end{center}
 \caption{One-dimensional energy spectra $E$ from LES with the Vreman
 model \cite{vrem:2004} with $C_v=0.07$ and anisotropic resolution of aspect ratio 32,
 compared with the equivalently
 filtered $|\bkappa|^{-5/3}$ Kolmogorov inertial energy spectra. Shown
 are spectra in both the fine and coarse directions.}
 \label{vre}
 \end{figure}

\begin{figure}[tp]
 \begin{center}
 \subfigure[Book]{\includegraphics
    [width=0.45\linewidth,bb=35 0 486 419,clip]{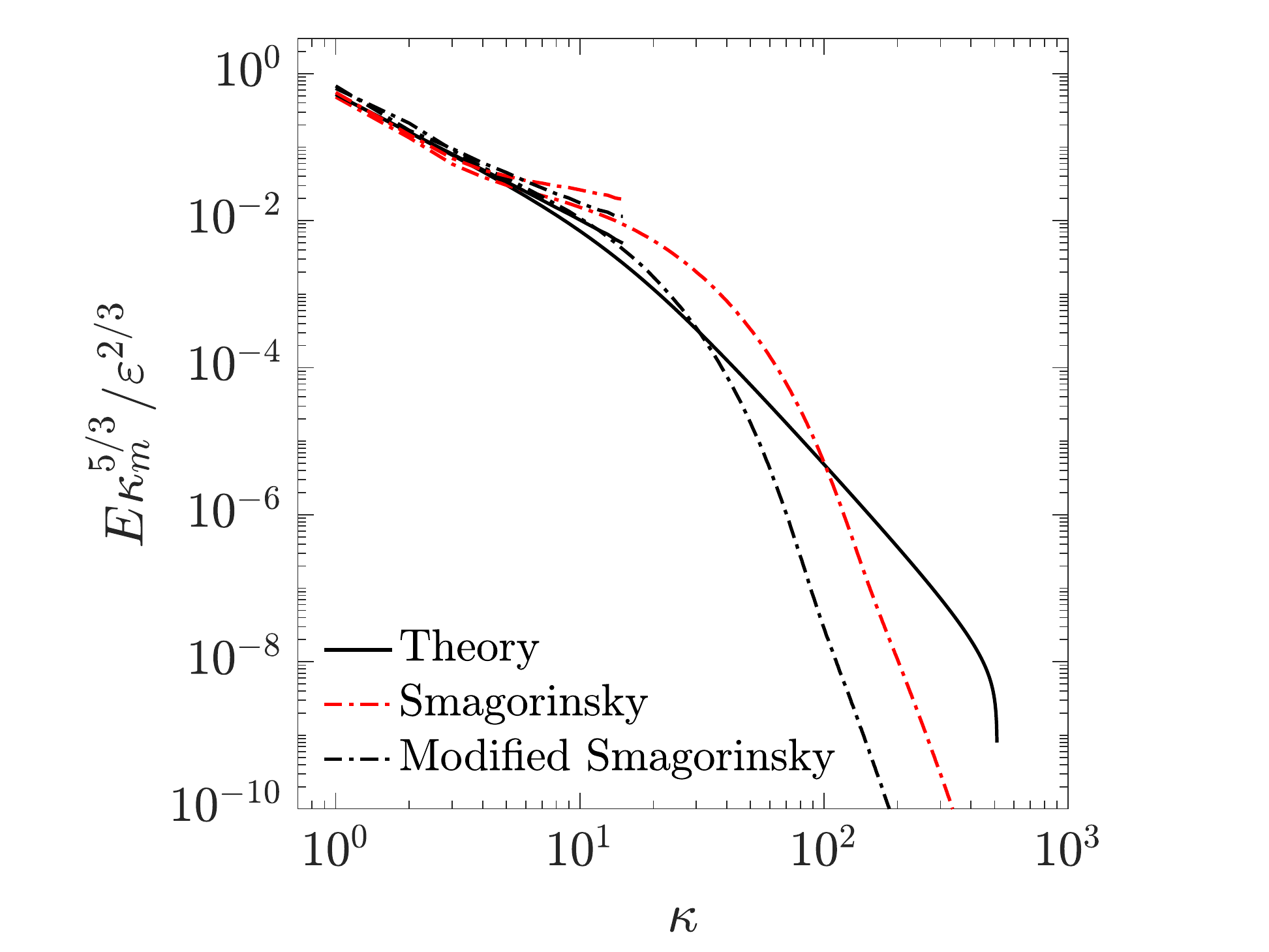}
    \label{asmag_b}}
 \subfigure[Pencil]{\includegraphics
    [width=0.45\linewidth,bb=35 0 486 419,clip]{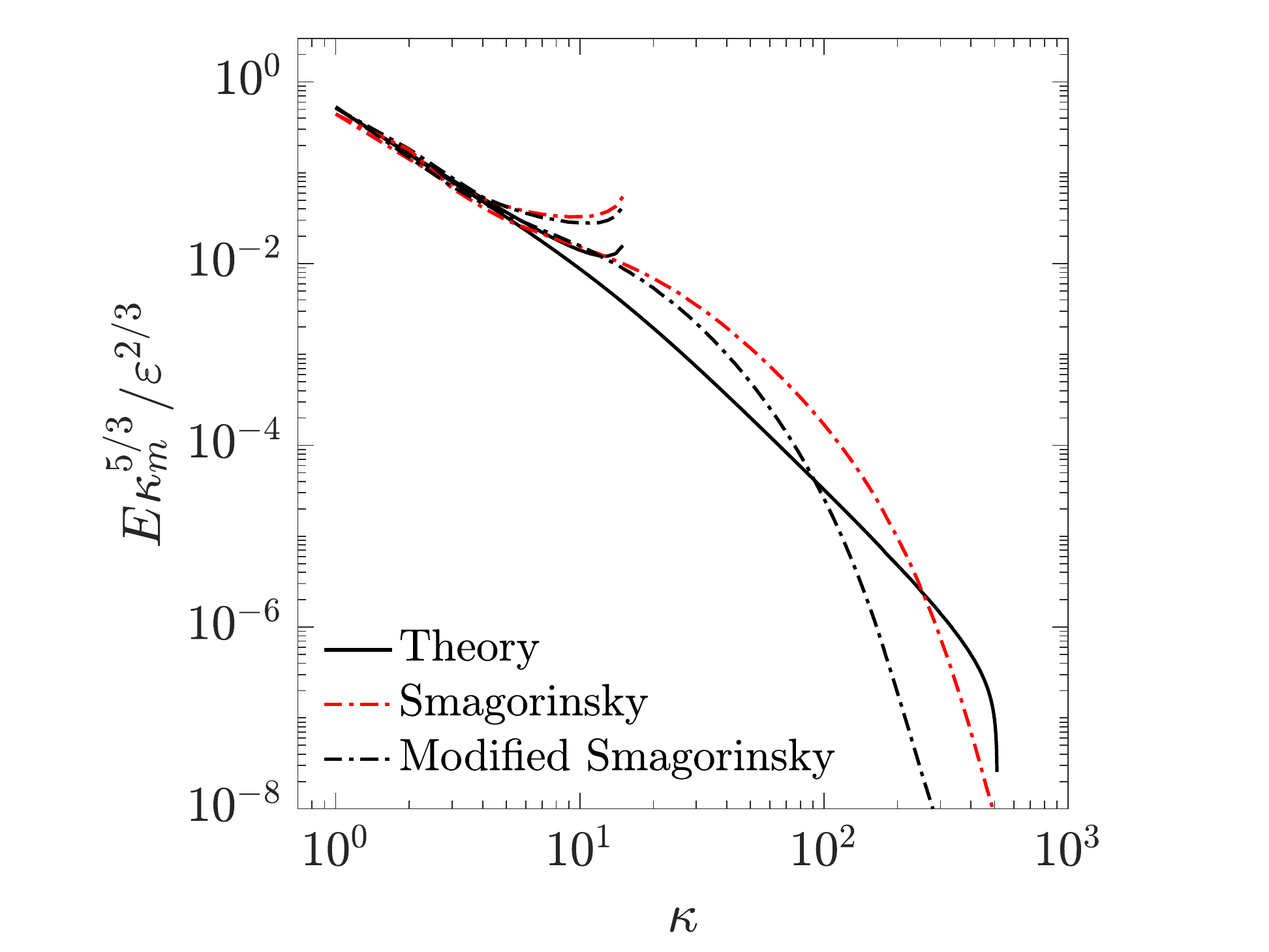}
    \label{asmag_p}} \\
 \end{center}
 \caption{One-dimensional energy spectra $E$ from LES with the basic
 Smagorinsky model and the modified Smagorinsky model of Scotti \cite{scott:1993},
 using anisotropic resolution with aspect ratio 32,
 compared with the equivalently
 filtered $|\bkappa|^{-5/3}$ Kolmogorov inertial energy spectra. Shown
 are spectra in both the fine and coarse directions.}
 \label{asmag}
 \end{figure}

As with the basic Smagorinsky model evaluated in Sec.~{\ref{sec:intro}}, models are
assessed based on their ability to predict the one-dimensional energy
\begin{figure}[tp]
 \begin{center}
 \subfigure[Book Fine]{\includegraphics
    [width=0.45\linewidth,bb=35 0 486 419,clip]{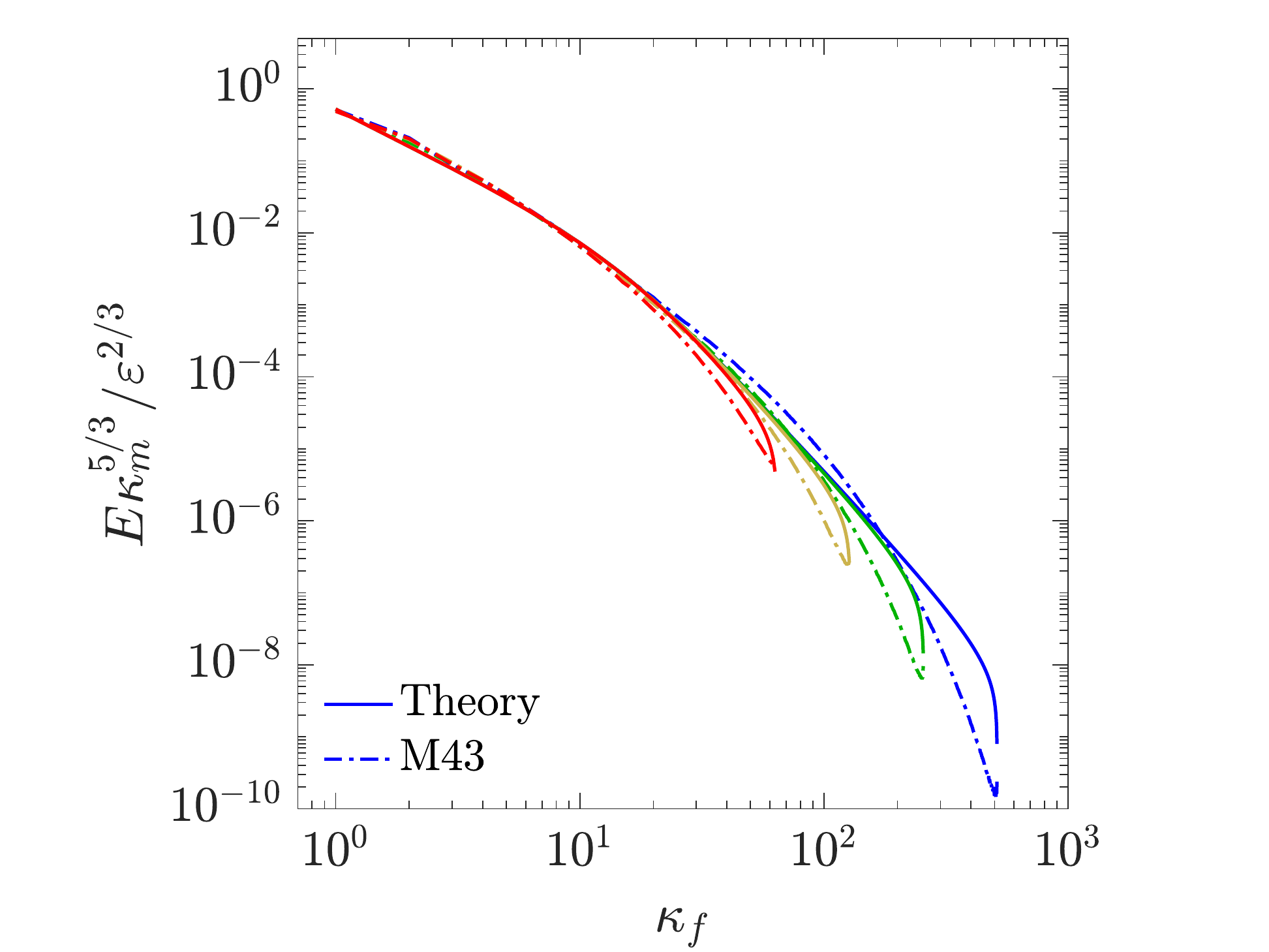}
    \label{m43_fbook}}
 \subfigure[Book Coarse]{\includegraphics
    [width=0.45\linewidth,bb=35 0 486 419,clip]{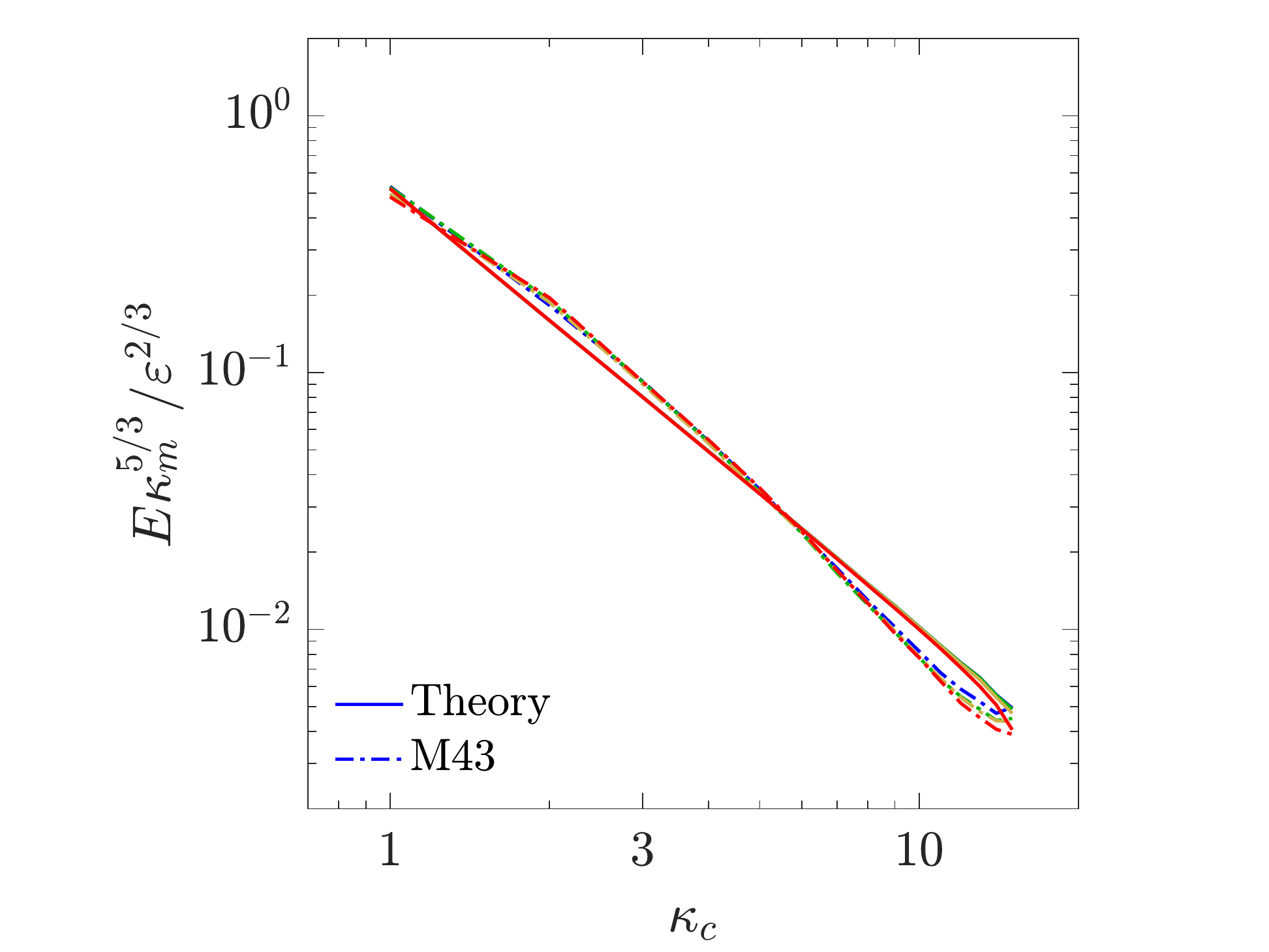}
    \label{m43_cbook}}\\
Cell aspect ratio: {\color{red} 4}, {\color{dy} 8},
 {\color{dg} 16}, {\color{blue} 32}\\
\subfigure[Pencil Fine]{\includegraphics
    [width=0.45\linewidth,bb=35 0 486 419,clip]{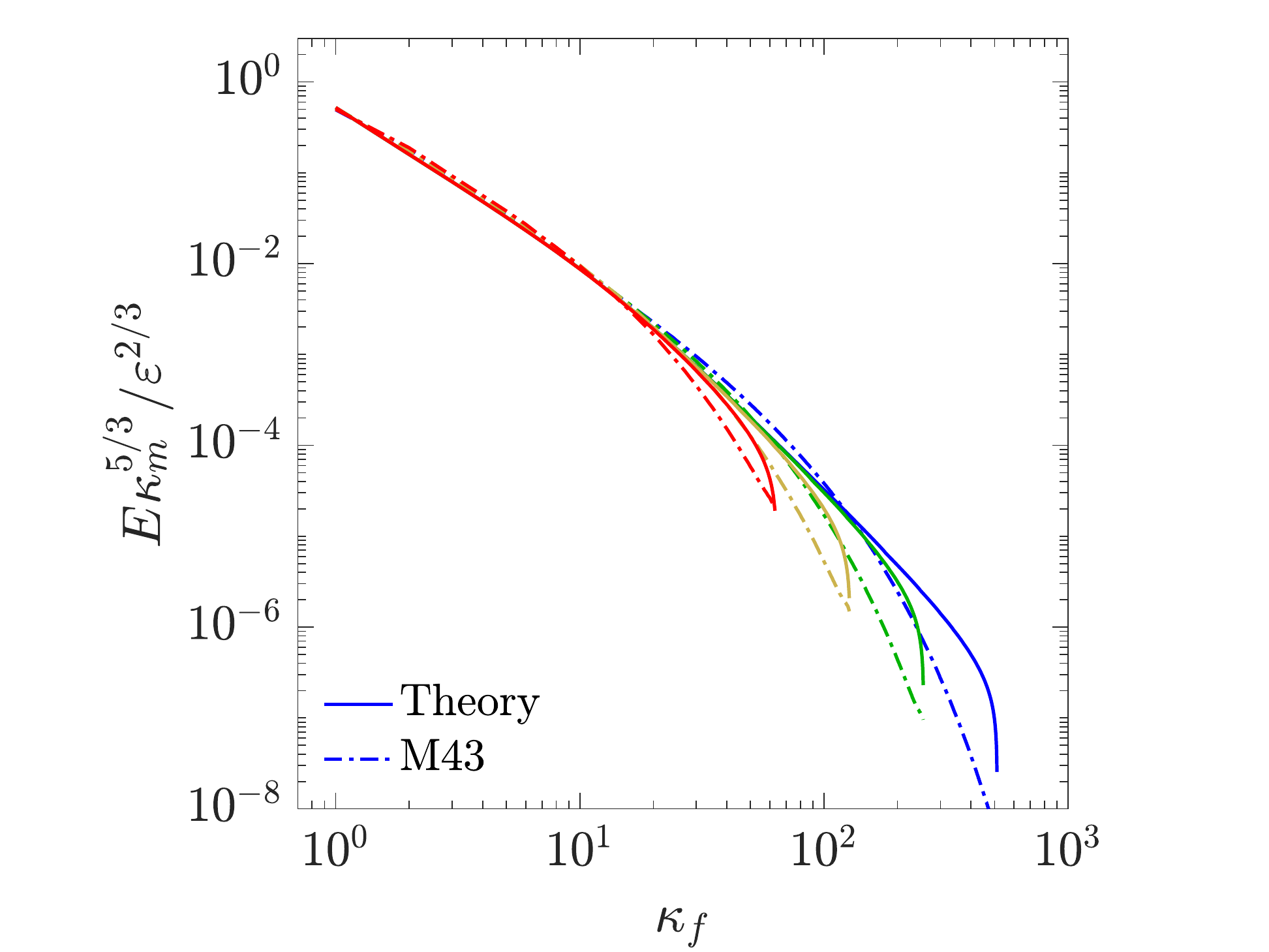}
    \label{m43_fpencil}}
 \subfigure[Pencil Coarse]{\includegraphics
    [width=0.45\linewidth,bb=35 0 486 419,clip]{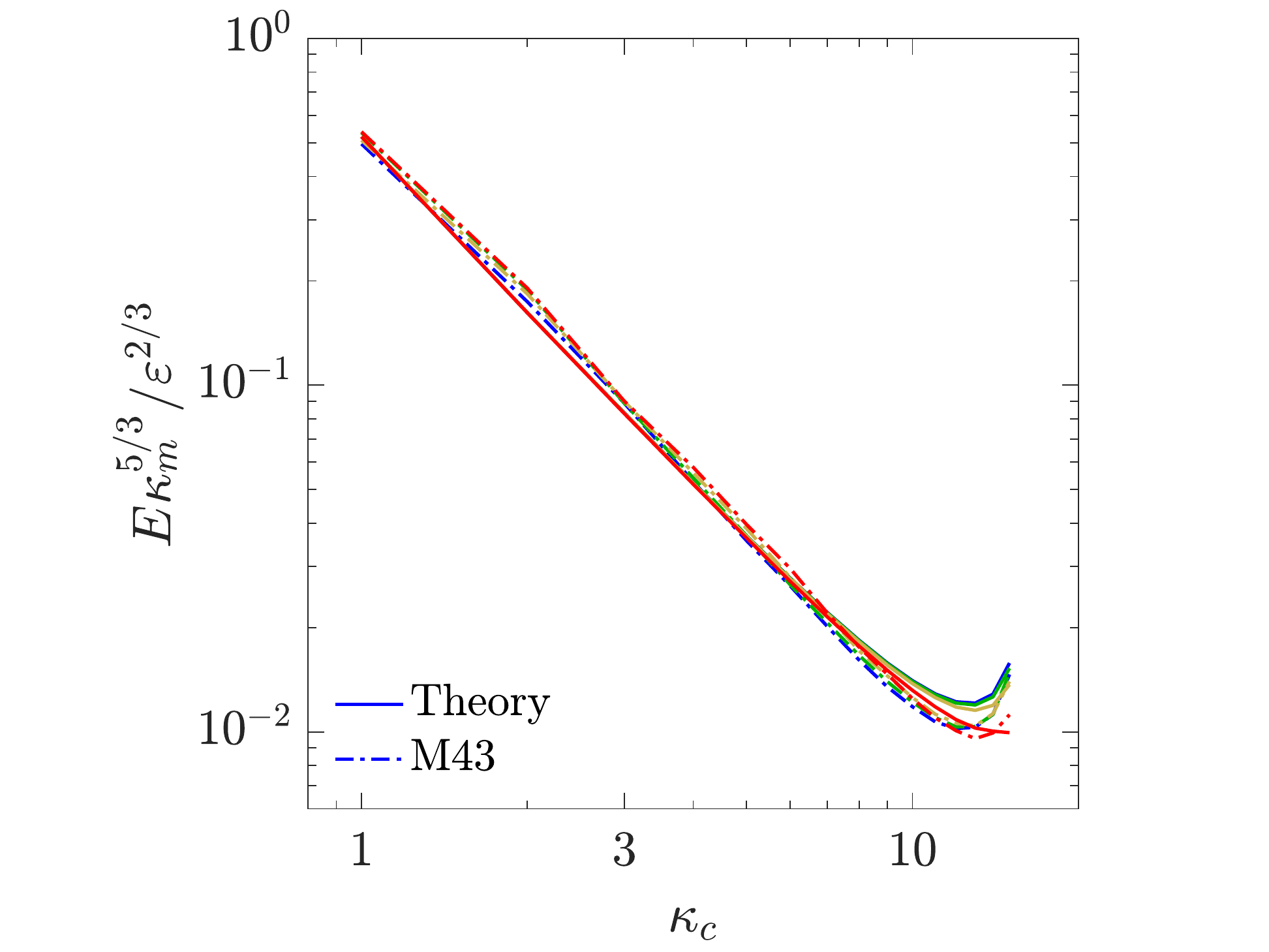}
    \label{m43_cpencil}} \\
\end{center}
\caption{One-dimensional energy spectra $E$ from LES with the M43
 model and anisotropic resolution, compared with the equivalently
 filtered $|\bkappa|^{-5/3}$ Kolmogorov inertial energy spectra.}
 \label{m43}
 \end{figure}
%
%
spectra when performing an LES of infinite Reynolds number forced
homogeneous isotropic turbulence with anisotropic resolution (see
Fig.~\ref{amd}-\ref{asmag}). Spectra obtained by filtering a
theoretical infinite Reynolds number inertial range spectrum is used
for comparison, to avoid the finite Reynolds number effects inherent
in DNS data.  This is critical when considering high aspect ratio
resolution because the models assume that the resolution scales in all
directions are in the inertial range. Comparisons between the LES and
the theoretical filtered spectrum are based on ellipsoidal sharp spectral
cut-off filters. That is the Fourier transform $\hat\cF(\bkappa)$ of
the filter kernel is given by:
\begin{equation}
  \hat\cF(\bkappa)=
  \begin{cases}
    1 & \text{if $\cM^2_{ij}\kappa_i\kappa_j < \pi^2$} \\
    0 & \text{otherwise}
  \end{cases}
  \label{EllipFilter}
\end{equation}
This filter excludes all wavenumbers except those in the domain
$\cD^e$ defined in (\ref{DefinecDe}). When simulating isotropic
turbulence with anisotropic resolution, such an ellipsoidally filtered
turbulence is the most meaningful target of the LES. However, the LES
numerical representation and the associated models, are based on a
Cartesian tensor product definition of the resolved turbulence
($\cD^c$ in Eq. \ref{DefinecDc}). Therefore, to obtain the spectra of
interest and compare to the theoretical spectra, the spectral filter
defined in (\ref{EllipFilter}) is also applied to the LES
solutions. Comparisons between LES and theory based on the Cartesian
tensor product definition of resolved scales used in the simulations
leads to the same conclusions regarding the fidelity of the models as
the comparisons reported here. This is also how the comparisons with
Smagorinsky LES in Fig.~\ref{smag} were performed.

One-dimensional energy
spectra are reported here instead of the three-dimensional spectra
more commonly used for isotropic turbulence, to reveal the differences
between the coarsely and finely resolved directions.

LES are performed with the $3/2$-dealiased pseudo-spectral code
\emph{PoongBack} \cite{LeeMoser2015}.  Negative viscosity forcing is performed over a
band of wavenumbers with magnitudes $|\bkappa|=(0.0,2.0]$.  All statistics
  are averaged over 10 fields spanning at least four eddy turnover times
  after being brought to a stationary-state.
  All the models considered here
  yield virtually identical results for isotropic resolution, \emph{e.g.} Fig. \ref{isores}.

We begin by performing the same book and pencil cell evaluations up to
aspect ratios of 32 as for the basic Smagorinksy model
(Fig.~\ref{smag}).  The AMD model results are shown in Fig.~\ref{amd}
with M43 shown in Figures \ref{m43}.  Both of these models perform
quite well in comparison to Smagorinsky (Fig.~\ref{smag}) with little
energy pile-up at the cutoff in the coarse direction.  In the AMD
model, there is a very small excess of energy in the coarse spectrum,
especially toward the cutoff  with pencil cells, which
rapidly saturates with increasing aspect ratio. With the M43
model, there is apparent over-dissipation in wavenumbers near the
cutoff primarily in fine directions for both cell types resulting in
reduced spectral energy near the cutoff.  This behavior appears to
increase with cell aspect ratio. The AMD model also produces spectra
that are too low near the cutoff in the fine direction, though the
character of the curves is more complex. Particularly, the spectra
roll off with a slope that is not monotonically increasing with wavenumber.

 To explore the effect of scale separation between the largest
 turbulent scales and the coarsest filter cutoff, we examine model
 performance as a function of the ratio of the coarsest cutoff
 wavenumber, $\kappa_{cc}$, to minimum resolved wavenumber,
 $\kappa_m$, with a fixed cell aspect ratio of 8 for book cells.  In
 the cases considered, the integral scale $L_{int}$ is given by
 $L_{int}\kappa_m\approx1.38$ , so varying $\kappa_{cc}/\kappa_m$
 similarly varies $L_{int}\kappa_{cc}$. Shown in figure \ref{Lratio}
 are spectra for $\kappa_{cc}/\kappa_m=2$, 4, 8 and 16. In addition to
 the M43 model in both its basic and low-k version and AMD, the basic
 Smagorinksy model is evaluated for reference (Fig.~\ref{smagL}).  The
 previously observed pileup of energy near the coarse cutoff for
 Smagorinsky increases relative to the total resolved energy as
 $\kappa_{cc}/\kappa_m$ is reduced.  Once again, the M43 models and
 AMD vastly out-perform Smagorinsky and yield nearly the theoretical
 spectra.  The low-k version of the M43 model is virtually
 indistinguishable from the basic M43, even for
 $\kappa_{cc}/\kappa_m=8$. It my be that the low-k correction will be important
 for smaller values of $\kappa_{cc}/\kappa_m$, though it is clearly
 not for the cases considered here.
 
Finally, we briefly discuss the Vreman model \cite{vrem:2004}
(Fig. \ref{vre}) and the Scotti modification of Smagorinsky
\cite{scott:1993} (Fig. \ref{asmag}).  For both models, the qualitative performance is similar to basic Smagorinsky, with 
energy pile up at the coarse direction cutoff and mid-range of the
fine direction, while the energy is low near the  
fine direction cutoff.  As the Vreman model was primarily designed to cause 
the model viscosity to vanish in laminar regions, poor results in the presence 
of anisotropic grids is not surprising. Scotti's modified Smagorinsky effectively increases the model constant in 
response to the cell aspect ratio.  Naturally, it cannot improve the basic anisotropic 
resolution behavior of the Smagorinsky model and can only reduce
coarse direction energy pile-ups by increasing the eddy diffusivity.
This is precisely the observed behavior.  However, especially for pencil cells, it appears the 
correction should be enhanced, as there continues to be excessive
energy at the cutoff in the coarse direction.\\*

\subsection{Dissipation anisotropy}
\label{sec:eval_diss}
\begin{figure}[tp]
 \begin{center}
 \subfigure[Book]{\includegraphics
    [width=0.45\linewidth,bb=58 0 483 420,clip]{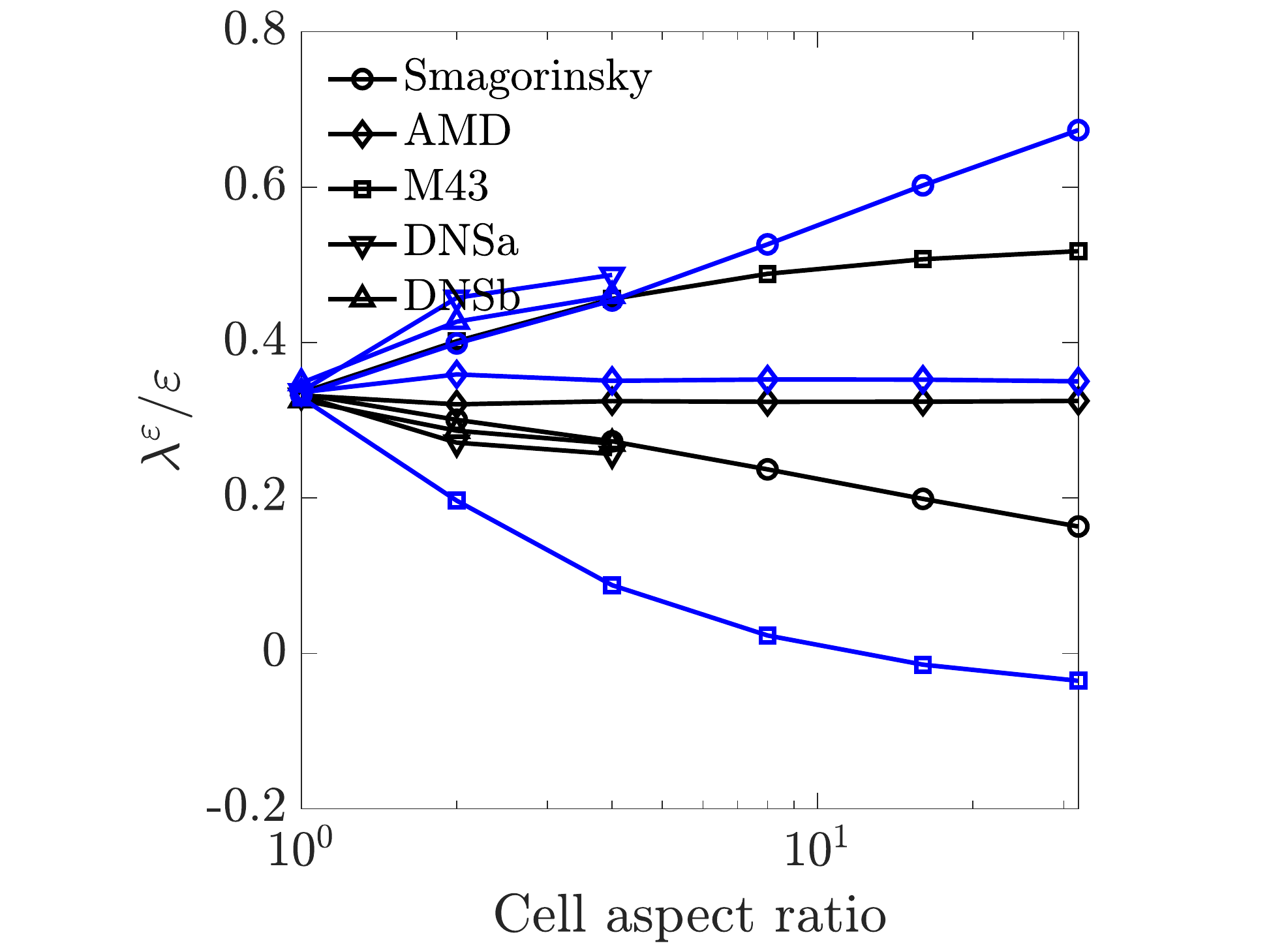}
    \label{eij_b}}
 \subfigure[Pencil]{\includegraphics
    [width=0.45\linewidth,bb=58 0 483 420,clip]{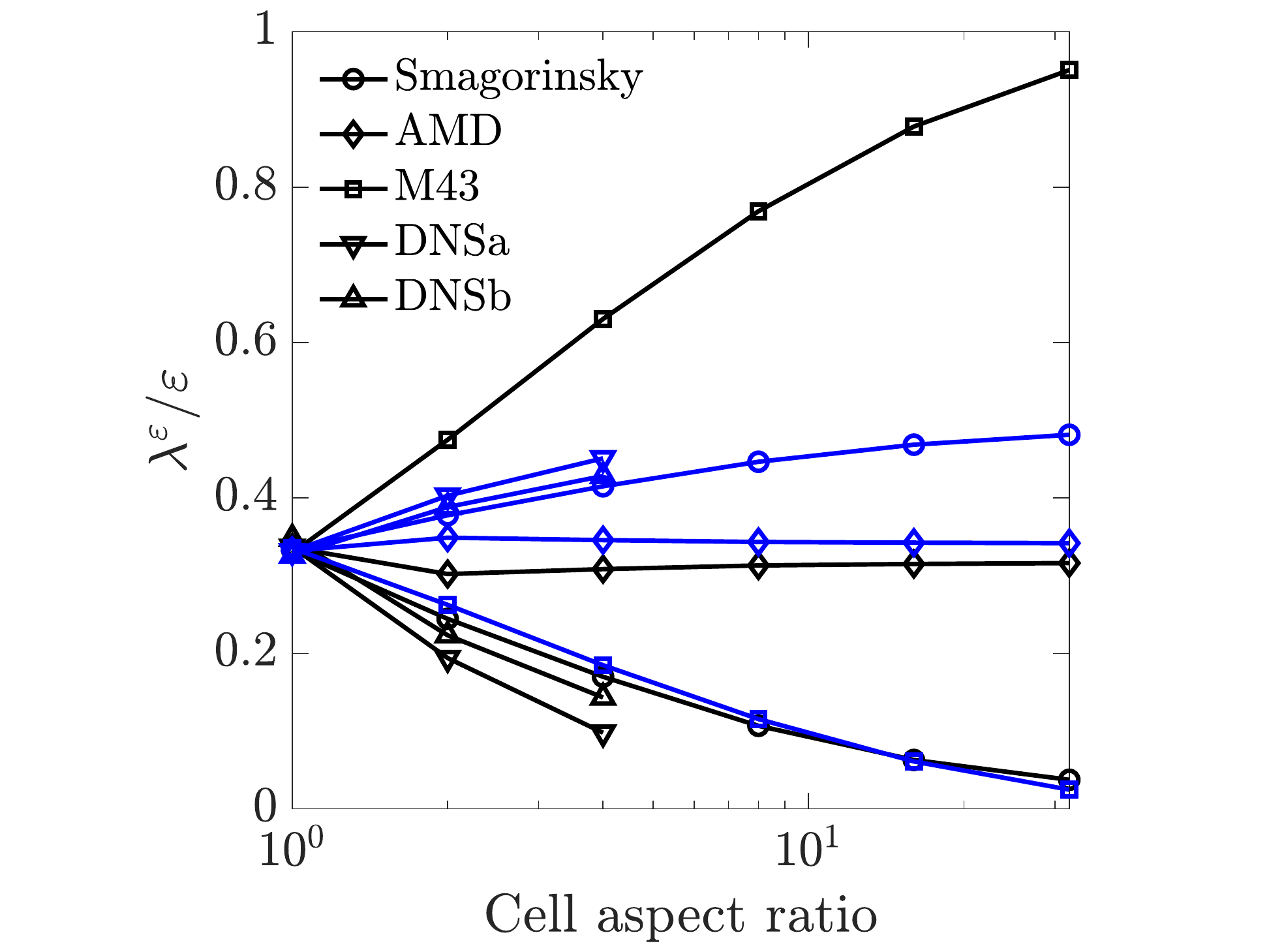}
    \label{eij_p}} \\
 \end{center}
 \caption{Normalized values of the $\varepsilon_{ij}$ eigenvalues as a
 function of cell aspect ratio in LES using the the basic Smagorinsky,
 AMD, and M43 models. Also shown are values from a filtered DNS
 with equivalent forcing and $Re_\lambda=205$ performed with $512^3$
 Fourier modes. The filters used on the DNS were consistent with
 LES resolution with $\kappa_{cc}/\kappa_m=32$ (DNSa) and
 $\kappa_{cc}/\kappa_m=16$ (DNSb). Black lines represent
 $\lambda^\epsilon_c$ and {\color{blue} blue} lines represent
 $\lambda^\epsilon_f$.}
\label{Feij}
 \end{figure}

Both the M43 model and the AMD model perform admirably
well on the anisotropic resolution cases studied here, despite the
prominent differences in their formulation. This raises the question as to
what they have in common that other models lack  that leads to the
good performance on these tests. An obvious candidate for this shared
feature is the anisotropic energy transfer to small scales
(\ref{eq:epsilonij}), which
measures the contributions of stress and velocity gradients in different directions
to the energy transfer.  In light of (\ref{e1}), it seems probable that this directional contribution to energy
transfer should have a direct impact on the one dimensional spectra
studied in Sec.~\ref{sec:eval_spectra}, and indeed the importance of this
energy transfer tensor was assumed in the formulation of the M43 model
(Sec.~\ref{sec:M43}). To test this hypothesis,
$\varepsilon_{ij}$ was computed from the LES presented in
Sec.~\ref{sec:eval_spectra} using the  M43, AMD, and Smagorinsky
models as a
function of resolution aspect ratio. Because the turbulence being
simulated is isotropic, the only source of anisotropy is the
resolution. This guarantees that $\varepsilon_{ij}$
has the same eigenvectors as $\cM$, regardless of model. Further,
because the LES are (isotropically) forced and stationary, the total
energy transfer $\varepsilon_{ii}$ is just the mean rate at which
energy is introduced by the forcing, which is also the same regardless
of model. Therefore, to compare $\varepsilon_{ij}$ for the different
models, it suffices to compare it's eigenvalues $\lambda^\varepsilon$
normalized by $\varepsilon_{ii}$. To this end, the normalized
eigenvalues associated with eigenvectors in the coarse and fine
resolution directions are shown as a function of aspect ratio in Fig. \ref{Feij}.

The results shown in Fig.~\ref{Feij} are surprising.  For both book
and pencil cells, the energy transfer with the M43 model is dominated
by gradients in the coarse direction(s).  For book cells, this trend
is so pronounced that for aspect ratio greater than 16, the
contribution of gradients in the fine direction become slightly
negative, that is representing net energy transfer from small to large
scales. The Smagorinsky results are exactly the opposite with
gradients in the fine direction(s) dominating the energy
transfer. When using the AMD model, $\varepsilon_{ij}$ is much
closer to isotropic, with the coarse and fine direction eigenvalues
nearly the same, resulting in normalized values of about a third.  Equivalently 
filtered $512^3$ DNS with identical forcing is also presented using
filters with the same 
$\kappa_{cc}/\kappa_m$ as the LES (32) and with $\kappa_{cc}/\kappa_m=16$, to 
reduce finite Reynolds number effects.  While the dissipation contribution of DNS 
does not match any model particularly well, it does most closely resemble 
that of the basic Smagorinsky model with the fine directions contributing the 
most.  The fact that both the AMD and M43 models perform well with 
anisotropic resolution, but produce strikingly different anisotropic 
characteristics of $\varepsilon_{ij}$ from both each other and DNS shows 
that reproducing this statistic is not necessary for capturing the anisotropy of the
resolved spectrum. This is curious, as it seems that $\varepsilon_{ij}$ 
should be relevant. It is also unfortunate, as it would be useful in formulating 
LES models to know statistical conditions that are necessary for good 
performance.

\subsection{Non-fluctuating eddy viscosity}
\label{sec:nofluc}
One of the interesting things about the M43 model presented here is
that the eddy viscosity does not fluctuate, instead it is considered a
mean quantity. This has some clear advantages that arise from the fact
that it makes analyzing the model much easier because one does not
need to consider correlations between a fluctuating eddy viscosity and
the fluctuating velocity gradient. However, with isotropic resolution,
the M43 model has a scalar eddy viscosity, and an LES with a
non-fluctuating scalar eddy viscosity would appear to be a DNS at some
low Reynolds number. That this is not so is a consequence of the
limited resolution used to represent the turbulent fluctuations. In the M43
model, the eddy viscosity was determined to ensure that when the model
is applied to turbulence with a Kolmogorov intertial range extending
up to the spectral resolution cutoff, that it will dissipate energy at
the rate $\varepsilon$. This means simply that the subgrid model
dissipation does not preclude a trucated $\kappa^{-5/3}$ spectrum as
the LES solution. Whether this is realized or not depends on the
details of the energy transfer among scales in the LES. This is
distinctly different from a DNS in which the resolution is sufficient
to capture the viscous roll off of the spectrum. With isotropic
resolution, the M43 eddy viscosity leads to an effective ``Kolmogorov
scale'' $\eta_e\approx 0.136\delta$ so that the effective
$k_c\eta_e\approx0.428$. This is much smaller than that considered
adequate resolution for DNS (e.g. $k_c\eta\ge1.5$ in \cite{DonzisSreeni2010}). 

\begin{figure}[tp]
 \begin{center}
  \subfigure[Without spherical filtering]{\includegraphics
     [width=0.45\linewidth,bb=35 0 486 419,clip]{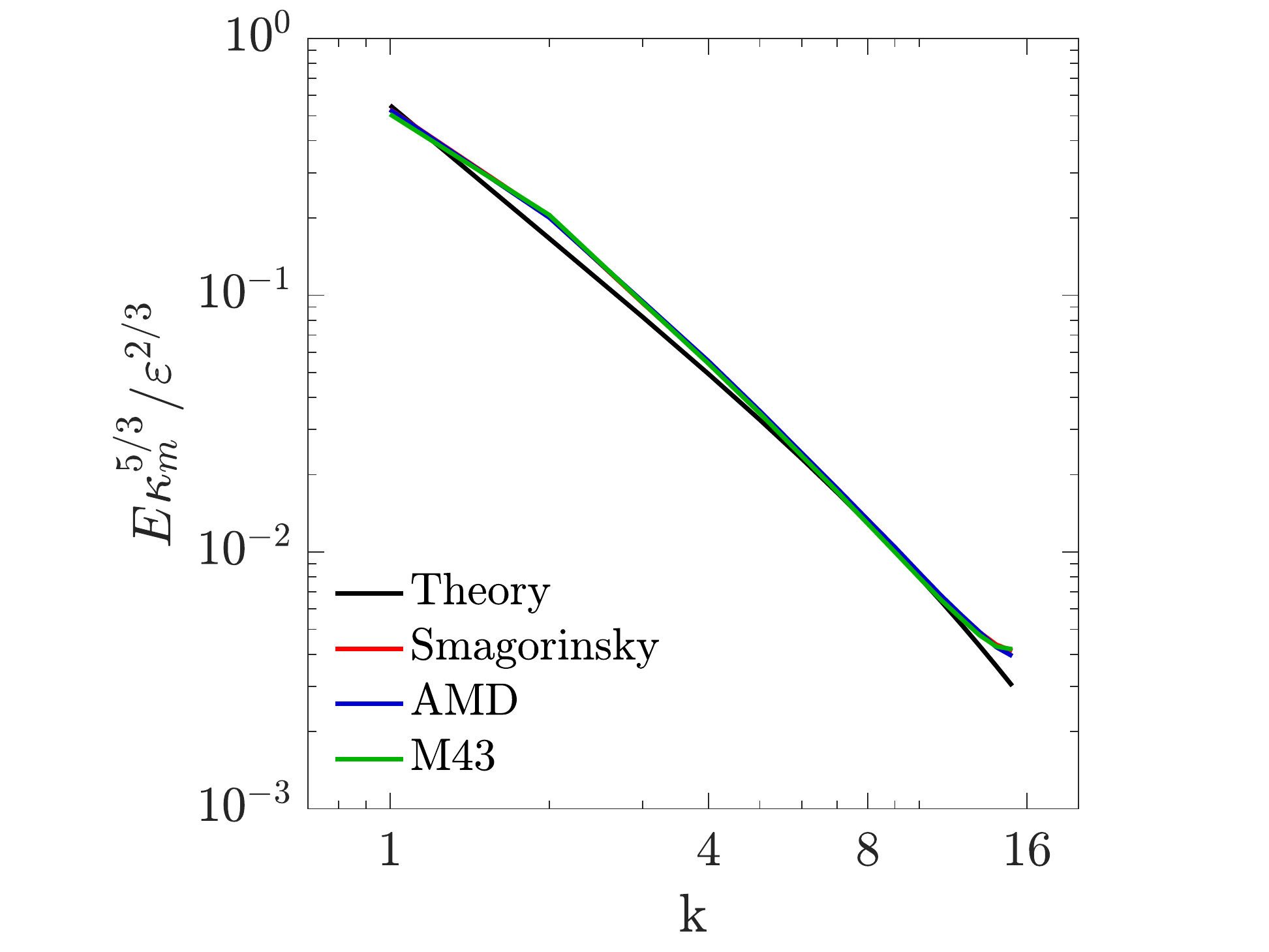}
     \label{isores_box}}
  \subfigure[With spherical filtering]{\includegraphics
     [width=0.45\linewidth,bb=35 0 486 419,clip]{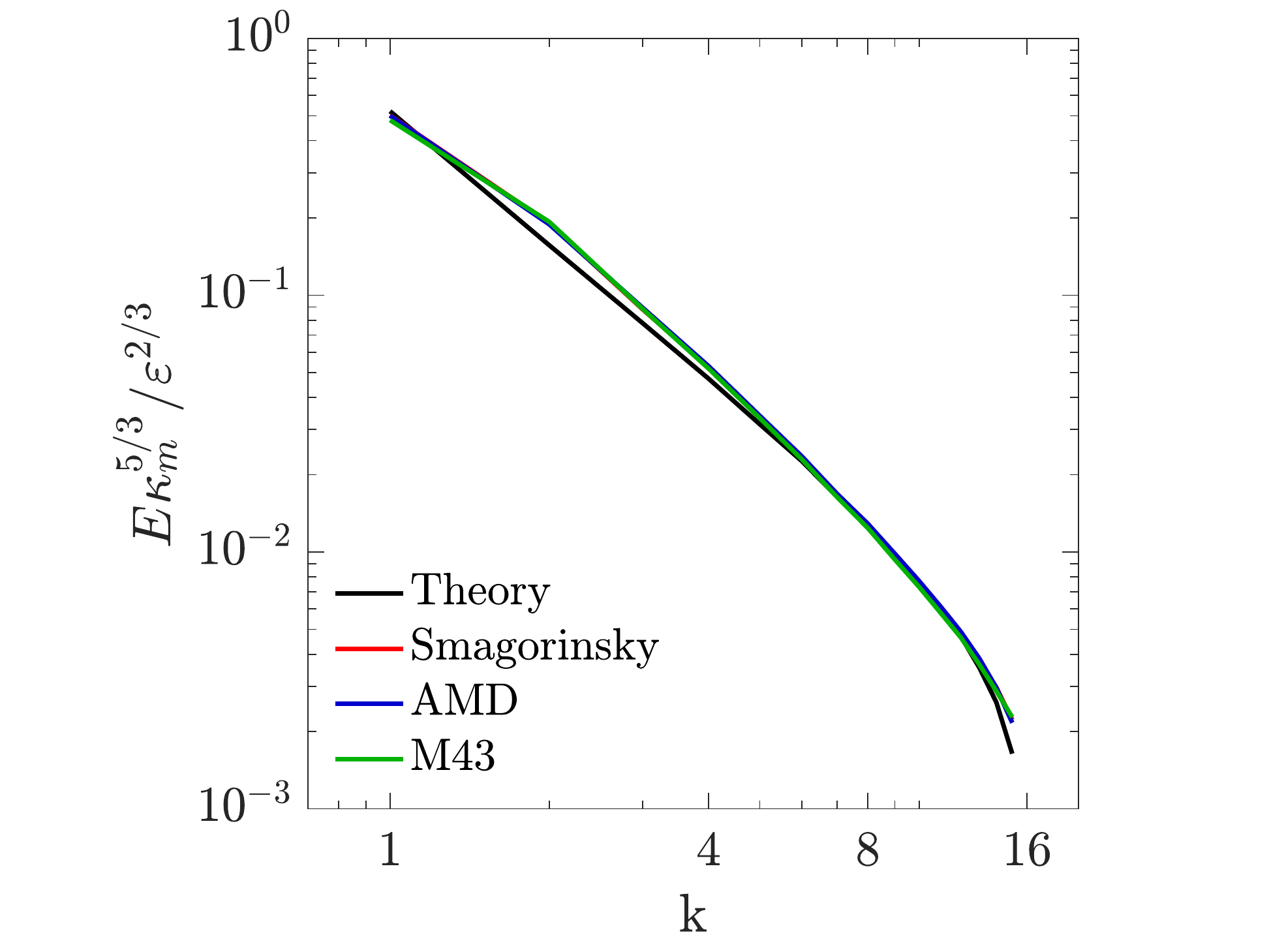}
     \label{isores_sphere}}\\
 \end{center}
 \caption{One-dimensional energy spectra $E$ from LES with isotropic
 resolution using the Smagorinsky, AMD and M43 models, compared with the equivalently
 filtered $|\bkappa|^{-5/3}$ Kolmogorov inertial energy spectra.
Shown are spectra (a) without a spherical cut-off filter and (b) with a
  spherical cut-off filter.
}
 \label{isores}
 \end{figure}
All the other models considered here use a fluctuating eddy viscosity,
and the question arises as to whether the fluctuations contribute to
the verasity of the model. It is at least plausible that a fluctuating
eddy viscosity could better represent the scale dependent transfer of
energy to unresolved scales, leading to a better representation of the
the truncated spectrum or perhaps other statistics. To investigate
this, consider the spectra from LES with isotropic resolution using
the Smagorinsky, AMD and M43 models shown in Fig.~\ref{isores}.
Shown are one-dimensional spectra computed with and without a spherical
truncation, which is the isotropic version of the ellipsoidal filter
used in the spectra shown in Sec.~\ref{sec:eval_spectra}. Without the spherical
filter, the spectra represent a tensor product spectral truncation
consistent with the numerical representation used in the LES. Notice
that with or without the spherical truncation, all three models
yield nearly identical specrtra, including an excess of energy
at the highest resolved wavenumber, compared to the filtered
theoretical spectra, which is manifested as an up-turn in the spectra
in the non-spherically-filtered case.

It appears then that, at least for the energy spectra, the
fluctuations of the eddy viscosity in the Smagorinsky and AMD models
do not improve the veracity of the model when applied to LES with
isotropic resolution. However, from Sec.~\ref{sec:eval_spectra}, the
fluctuating eddy viscosity does allow the AMD model to treat
anisotropic resolution without resorting to an anisotropic eddy
viscosity, though this is not true for the Smagorinsky model.

\section{Conclusions}
\label{sec:conclusion}

Application of LES in complex geometries, or even just wall bounded
flows, will generally involve grids with anisotropic resolution. Even
when the turbulence is isotropic, anisotropic LES resolution results
in resolved and unresolved turbulence that are anisotropic, violating
the isotropic unresolved scales assumption underlying many subgrid
models. Indeed it was shown (sections~\ref{sec:aniso} and
\ref{sec:eval_spectra}) that commonly used LES models perform poorly
on LES of isotropic turbulence with anisotropic resolution. The contribution
of the unresolved scales to the mean Reynolds stress is only mildly
anisotropic. In contrast, as expected, the resolved velocity gradients
are strongly anisotropic, as evidenced by the anisotropy of the
quadratic product of the resolved gradients. Because the resolved
gradients are strongly anisotropic, and the subgrid contribution to
the Reynolds stress is only mildly so, one cannot expect a scalar eddy
viscosity to correctly represent the latter in general.  A potential method to 
correct this issue is to construct eddy viscosities to be uncorrelated with the 
gradient fluctuations so that its contribution to the mean stress only acts through the mean gradients \cite{urib:2010}. 
This is another sense in which LES subgrid models are ill-suited to modeling 
the unresolved Reynolds stress, as first observed by Jimenez \& Moser \cite{moser:2000}.

Treatment of resolution anisotropy in LES has been largely
neglected. In the few examples where resolution anisotropy has been
considered in formulating LES models, it has not been central to the
formulation, and the result is poor performance on the
simulations of isotropic turbulence with anisotropic resolution
performed here. The typical poor performance of these and other standard
models (e.g. Smagorinsky) is a pile up of spectral energy near the
resolution cut-off in coarse directions and/or too rapid a spectral
roll-off in the fine directions.  The one exception is the AMD model
introduced in \cite{roze:2015}, which uses a scalar eddy viscosity 
which incorporates resolution anisotropy and performs well on our
tests.

In this paper, we introduce a new subgrid model formulation to treat
resolution anisotropy based on a second rank tensor eddy
viscosity. In this M43 model, the anisotropy of the eddy viscosity is
determined by the 4/3 power of the resolution tensor $\cM$, thus the
name. In it’s simplest form, the eddy viscosity is just
$C\varepsilon^{1/3}\cM^{4/3}$, where $\varepsilon$ is the mean rate of
kinetic energy dissipation and $C$ is a dimensionless scalar function
of the invariants of $\cM$. This model performs similarly to
the AMD model in producing the correct anisotropic spectra in our
simulations of isotropic turbulence with anisotropic
resolution. However, the AMD and M43 models could not be more
different, and considering these differences and the fact that they do
not result in significant performance differences yields a number of
insights into what is important about an LES model, at least for
representing resolution anisotropy.

First, consider that the only flow-dependence in the M43 model is the
mean rate of kinetic energy dissipation. This means that the M43 eddy
viscosity is independent of time for stationary flows, is independent
of homogeneous spatial directions, and in general only varies on the
scale of mean variations, not on the scale of the resolved
fluctuations. In virtually every other LES model, including AMD, the
eddy viscosity is spatially varying due to its dependence on local
instantaneous flow characteristics (e.g. the velocity gradients). The
good performance of the M43 model shows that it is not necessary for
an LES eddy viscosity to fluctuate or depend on fluctuating resolved
quantities. Though it is necessary that such an eddy viscosity produce
the correct dissipation, as has been widely understood. Indeed, one
interpretation of the dynamic Smagorinsky model \cite{germ:1991}, is that
the dynamic procedure serves to ensure that the dissipation rate is
consistent \cite{moser:2000}. The current results suggest that a dynamic model
could be formulated in which the mean dissipation rate was the
quantity that needs to be determined dynamically, and the dynamic process acts on averaged quantities.

Second, the mechanisms by which the AMD and M43 models represent
resolution anisotropy are completely different. For M43, the
representation is direct with the anisotropy of a tensor eddy
viscosity determined directly from the anisotropy of the resolution
tensor (\ref{M43}). In contrast, the AMD model has a scalar eddy
viscosity so that the anisotropic characteristics of the model arise
entirely from the anisotropic correlation of the fluctuating eddy
viscosity with the resolved velocity gradients. It is not clear why
this should work so well for anisotropic resolution in the AMD model. 
The AMD modeling ansatz of setting the eddy viscosity to an
estimate of the minimum required to dissipate the variance of the
velocity gradient at the rate that it is produced does not appear to
speak to this correlation characteristic. So while the M43 model
developed here is constructed specifically to perform well with
anisotropic resolution, the good AMD performance just arose in a model
developed based on other considerations. For future model refinement
and development, it would be useful to determine the characteristics
of AMD that lead to this good performance.

Also of interest is the fact that the {\it a posteriori} anisotropy of
the energy transfer tensor $\varepsilon_{ij}$ as defined in (\ref{eq:epsilonij}) is
essentially different between LES performed with M43 and AMD models, and
that these are also essentially different from $\varepsilon_{ij}$ in
filtered DNS (section~\ref{sec:eval_diss}). Since this tensor characterizes the
contributions of gradients in different directions to the transfer of
energy to the unresolved scales, it would seem to be critical to the
dynamics of the resolved scales. But the results in
section~\ref{sec:eval_spectra} show that correctly representing this quantity in
an LES is not necessary for good performance. This raises the question
of what statistical characteristics of the subgrid model are necessary
for good performance with anisotropic resolution. Knowing this, one
could design models that have these characteristics.

Finally, we note that the M43 model proposed here has some
features to recommend it beyond the relatively good performance for
LES with anisotropic resolution.
%
%
One is that the model, in theory, has no adjustable constants, or
rather, the model constant is determined in terms of the Kolmogorov
constant. The constant appearing in (\ref{M43}), which is a function
of $\cM$ can be determined from the Kolmogorov inertial range
spectrum, and characteristics of the numerical derivative operators
and filter type (Appendix A). Unlike the theoretical determination of
the Smagorinsky constant as in \cite{Lilly:1967}, $C(\cM)$ determined
in this way does result in good performance. Presumably this is
because in the M43 model, there are no fluctuations in the eddy
viscosity, so that the correlation of the eddy viscosity fluctuations
and the velocity gradient fluctuations do not contribute. 

Another useful feature is that the model is
formulated directly in terms of the mean rate of energy transfer to
the unresolved scales ($\varepsilon$). This might appear to be a
liability, since $\varepsilon$ is not generally known {\it a
priori}. However, because it is a well-defined scalar statistical
quantity that at high Reynolds number is independent of the scale at
which it is determined, it can be naturally found
dynamically. Alternatively, if one is carrying equations for the
dissipation, as for example in a hybrid RANS/LES formulation, that
dissipation can be used \cite{haer:2018}.

The lack of eddy viscosity fluctuations in the M43 model does not
limit the accuracy of the model in reproducing the energy spectra. The
presence of eddy viscosity fluctuations appears to be of no utility in
the Smagorinsky model. In the AMD model, on the other hand, they allow
resolution anisotropy to be treated with a scalar rather than the
tensor eddy viscosity used in M43. These results indicate that the
fluctuations in the eddy viscosity that commonly occur in subgrid
models are not necessary for good performance in an LES.

\section*{Appendix A: M43 coefficient}
\label{sec:App_A}

Let $\mathbb{G}$ be the normalized version of $\mathcal{G}$ as defined
in (\ref{G}); that is,
\begin{equation}
\mathbb{G}_{ijkl}=\mathcal{G}_{ijkl}\varepsilon^{-2/3}\delta^{4/3}
\label{G}
\end{equation}
where $\delta$ is the minimum eigenvalue of the resolution tensor,
$\mathcal{M}$. The normalized tensor (\ref{G}) then depends only on
the scaled resolution tensor $\hat{\cM}=\cM/\delta$.  
For isotropic turbulence, we can evaluate (\ref{G}) numerically for a wide range of 
resolution anisotropies by assuming a Kolmogorov inertial range
resulting in the scaled version of (\ref{Gijkl}). This calculation of $\mathbb{G}$ can account for the use of numerical
approximations of derivatives in computing the velocity gradients in
an LES, and the implicit or explicit filter defining the LES. This is
accomplished by introducing the effective wavenumber $\hat{\bm\kappa}(\bm\kappa)$
for the numerical approximation of the first derivatives into
(\ref{Gijkl}), and applying the homogeneous filter operator to the
integrand, yielding
\begin{equation}
\mathbb{G}_{ijkl}
= 
\frac{C_k\delta^{4/3}}{4\pi}\int_\cD\hat\kappa_k\hat\kappa_l\cF^2(\bm{\kappa})|\bkappa|^{-11/3}\bigg{(}\delta_{ij}-\frac{\kappa_i\kappa_j}{|\bkappa|^2}\bigg{)}{}d\bm{\kappa},
\label{Gijklnum}
\end{equation}
where $\cF$ is the filter operator. For example, when a finite volume
(box) filter is used, $\cF(\bm\kappa)=\Pi_{i=1}^3\mathrm{sinc}(\kappa_i\lambda_i^\cM/2)$.
For the LES performed here using the spectral numerical method in
\emph{PoongBack}, $\hat{\bm{\kappa}}=\bm{\kappa}$ and $\cF=1$. Consistent
with the numerical representation in \emph{PoongBack}, the domain of
integration is the Cartesian domain $\cD^c$ defined in
(\ref{DefinecDc}). The minimum wavenumber is $\kappa_m=2\pi/L$, where $L$
is the domain size, and the cutoff wavenumbers in each of the
principle directions of $\cM$ are
$\kappa_{c\alpha}=\pi/\lambda^\cM_\alpha$, where as in section~\ref{sec:M43}, $\lambda^\cM_i$
is the $\alpha$th eigenvalue of $\cM$. Reexpressing (\ref{DefinecDc})
in these terms yields:
\begin{equation}
\cD=\cD^c=\{\bm{\kappa}
| \kappa_m\le |\bkappa\cdot\bm{\phi}^\cM_\alpha|< \kappa_{c\alpha};\; \alpha=1,2,3\},
\end{equation}
where $\bm{\phi}^\cM_\alpha$ is the unit eigenvector of $\cM$ associated
with $\lambda^\cM_\alpha$. In a more general setting (other than turbulence
in a periodic box), the large scale $L$ would be proportional to the
integral scale. For any $\mathcal{M}$, when expressed in the Cartesian
basis defined by the the eigenvectors of $\cM$, (\ref{Gijklnum})
is non-zero only when each index value is repeated.

For a given resolution anisotropy, the model coefficient
$C(\hat{\cM})$ in (\ref{M43}) is determined to ensure that the eddy
viscosity tensor will produce the specified dissipation, and likewise
for $C^*(\hat{\cM}^*)$ for the low-k version of the model. To this end
the eddy viscosity model (\ref{M43}) is substituted into the trace of
the expression for the dissipation tensor (\ref{e1}). Since
$\varepsilon_{ii}=\varepsilon$, the result can be solved for $C$, yielding
\begin{equation}
C(\hat{\cM})=\Big{(}\hat{\mathcal{M}}_{jk}^{4/3}\mathbb{G}_{lljk} + \hat{\mathcal{M}}_{jk}^{4/3}\mathbb{G}_{lljk} + \hat{\mathcal{M}}_{lk}^{4/3}(\mathbb{G}_{ljjk}+\mathbb{G}_{ljjk})\Big{)}^{-1}
\label{CM}
\end{equation}
and the same expression is used to evaluate $C^*(\hat{\cM^*})$, by
substituting $\hat{\cM^*}$ for $\hat{\cM}$ in (\ref{CM}). There
remains only a single free constant, the Kolmogorov constant $C_k$,
that enters the model of $\cG$ in (\ref{Gijkl}). The overall model
constant ($C^\circ_\cM$ below) is calculated by considering the
isotropic resolution case in the limit where
$\kappa_c\gg\kappa_m$. In this case $\hat{\cM}=\cI$, the identity, and
(\ref{CM}) simplifies to
\begin{equation}
C_\cM^\circ=C(\cI)=\frac{4\pi\delta^{4/3}}{C_k}\left(\int_\cD
2|\bkappa|^{-5/3}\,d\bkappa\right)^{-1}
=\frac{2}{C_k\pi^{1/3}}\left(8\int_{\tilde\cD}|\tilde{\bkappa}|^{-5/3}\,d\tilde{\bkappa}\right)^{-1}\approx\frac{0.1106}{C_k}
\label{C0value}
\end{equation}
where the domain $\tilde\cD$ is the unit cube with one vertex at the
origin. Taking $C_k\approx 1.58$ as found in \cite{DonzisSreeni2010}
yields $C(\cI)\approx 0.070$, which is used here. This is about 1.5\%
lager than the value found by calibrating an LES with isotropic
resolution to match the filtered theoretical spectrum as closely as
possible, which yields only slight improvements in the spectrum. It
appears that 1.5\% is well within the uncertainty in the Kolmogorov
constant.

Since $C$ is
a scalar, it can only depend on the eigenvalues of $\hat{\cM}$, and
since by construction one of those eigenvalues is one, this is a two
dimensional function. It can be evaluated for a wide range of
resolution anisotropies characterized by the ratio of the
eigenvalues of $\cM$ to its minimum eigenvalue. A fit of this function
is described below.


\begin{table}[t]
\begin{tabular}{lSSS}
\hline
\noalign{\vskip 2pt}
\hline
\noalign{\vskip 2pt}
&\multicolumn{1}{c}{M43}&\multicolumn{1}{c}{M43 low-k}\\
\noalign{\vskip 2pt}
\hline
\noalign{\vskip 2pt}
$C^\circ_\cM$&0.070&0.070\\
\noalign{\vskip 2pt}
$c_{00}$&0.9091&             0.9091\\                               
$c_{10}$&0.2733&             0.2738\\                              
$c_{01}$&0.01989&            0.01848\\                               
$c_{20}$&-0.03121&           -0.03163\\                              
$c_{11}$&-0.1472&            -0.1472\\                             
$c_{02}$&0.01996&             0.01881\\  
$c_{30}$&-0.003754&            -0.003652\\                            
$c_{21}$&0.02011&            0.02012\\                               
$c_{12}$&-0.002831&           -0.00297\\                             
$c_{03}$&0.02067&            0.02022\\                               
$c_{40}$&0.0006687&             0.0006608\\
$c_{31}$&-0.0006634&          -0.0006655\\        
$c_{22}$&0.0011555&          0.001164\\                         
$c_{13}$&0.001673&          0.001663\\                            
$c_{04}$&0.003499&           0.003445\\
\noalign{\vskip 2pt}
\hline
\noalign{\vskip 2pt}
\hline
\end{tabular}
\caption{Values of the fitting coefficients in (\ref{fits}) based on
$\cG$ computed with $L/\lambda^\cM_{max}=64$ for the M43 and M43 low-k
models. The value of $C^\circ_\cM$ was determined from (\ref{C0value}).}
\label{tab:coefs}
\end{table}
   
\begin{figure}[t]
 \begin{center}
 \includegraphics[width=0.45\linewidth]{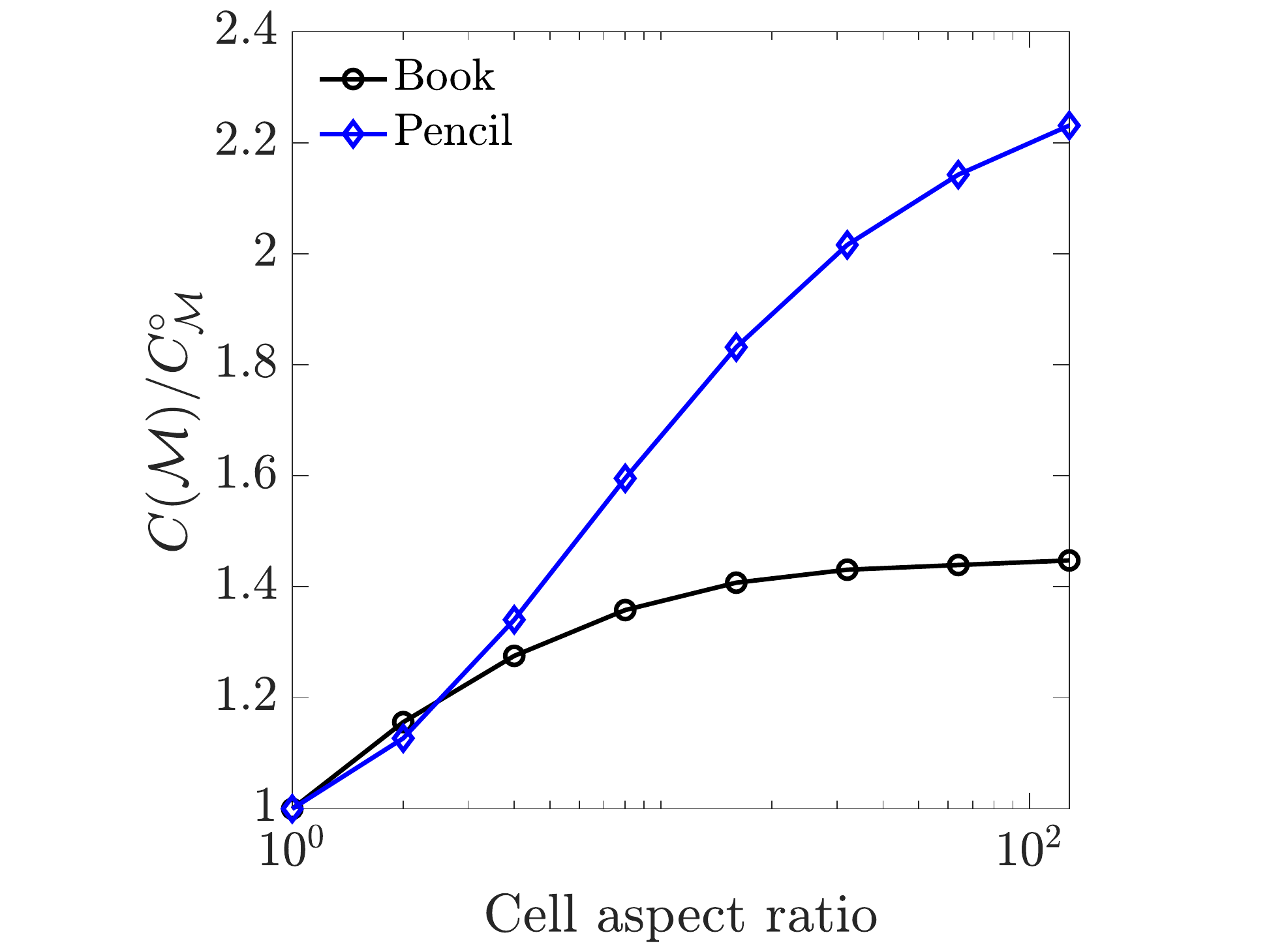}
 \end{center}
 \caption{Normalized coefficients for the basic M43 model as a function of cell aspect ratio using 
 (\ref{fits}).  All other cell types fall in between book and pencil limiting cases.}
 \label{Coefs}
 \end{figure}

Without loss of generality, let $\lambda^\cM_3$ be the smallest
eigenvalue of $\cM$ and $\lambda^\cM_1$ the largest. Then $C(\cM)$
depends only on $\lambda^{\hat\cM}_1>\lambda^{\hat\cM}_2\ge 1$. Let
$r^2=(\lambda_1^{\hat\cM})^2 + (\lambda_2^{\hat\cM})^2$ and let
$\theta=\cos^{-1}(\lambda_1^{\hat\cM}/r)$, where $0\le\theta\le\pi/4$.  The function $C(\cM)$ is
then fit as a quadratic function of $x=\ln(r)$ and
$y=\ln(\sin(2\theta))$. That is:
\begin{equation}
C(\cM)\approx C^\circ_\cM\sum_{i=0}^4\sum_{j=0}^{i-4} c_{ij}x^iy^j
\label{fits}
\end{equation}
Here the values of $c_{ij}$ are normalized so that for isotropic
resolution ($r=\sqrt{2}$, $\theta=\pi/4$), the sum in (\ref{fits}) is
1, and then $C^\circ_\cM=C(\cI)$ as determined above. For the M43
simulations performed here, fits including aspect ratios up to 128
were performed for spectral numerics with $\cG$ computed with
$L/\lambda_1^\cM=64$. This is large enough for the dependence on $L$
to be weak (see figure~\ref{g_aniso}). The resulting values of
$c_{ij}$ are given in table~\ref{tab:coefs}, and the values of
$C(\cM)/C^\circ_\cM$ for book and pencil resolution are plotted in
figure~\ref{Coefs}.

\section*{Appendix B: Modified AMD model}
\label{sec:App_B}

The AMD model was introduced by Rozema \textit{et al.}
\cite{roze:2015}. It was evaluated in section~\ref{sec:eval} and found
to perform remarkably well on isotropic turbulence LES with
anisotropic resolution. However, there is a technical detail in
the formulation of the model that needs be addressed, which is
discussed briefly here.

The expression $(\delta x_i \partial _i v_j)(\delta x_i \partial_i v_j)$
is introduced in equation (18) of \cite{roze:2015}, and similar
expressions are used throughout the subsequent development. Here
$\delta x_i$ is the size of the rectangular filter box in the $i$
direction (the grid size in the $i$ direction of a Cartesian grid) and
$\partial_i v_j$ is the velocity gradient tensor. The expression
$\delta x_i\partial_i$ is described in \cite{roze:2015} as the scaled
gradient operator. Throughout the
paper, Cartesian tensor notation and the Einstein summation convention
employed. However the expression above and those like it throughout
the paper are not valid Cartesian tensor expressions because the index
$i$ appears four times. As a consequence, the meaning of the expression
is ambiguous, but from context and by comparison to the QR model
development, it is clear that what is meant is
\begin{equation}
  (\delta x_i \partial _i v_j)(\delta x_i\partial_i v_j)=\sum_{i=1}^3(\delta x_i
  \partial _i v_j)(\delta x_i\partial_i v_j).
  \label{ScaledGradient}
\end{equation}
The reason this cannot be expressed in a valid Cartesian tensor
expression is that the directional filter sizes $\delta x_i$, do not
make up a tensorially valid vector. In essence, (\ref{ScaledGradient})
can only make sense when the Cartesian basis vectors are normal to the
faces of the filter box.

This difficulty can be addressed by observing that the $\delta x_i$
are actually the eigenvalues of the resolution tensor $\cM$ introduced
in section~\ref{sec:M43}, and in the case of rectangular filter boxes,
the eigenvectors of $\cM$ are normal to the faces of the box. A
tensorially consistent representation of the scaled gradient is then
$\cM_{ik}\partial_k$, and the expression in (\ref{ScaledGradient}) is
written
\begin{equation}
  \sum_{i=1}^3(\delta x_i \partial _i v_j)(\delta x_i\partial_i v_j) =
  (\cM_{ik}\partial_k v_j)(\cM_{il}\partial_l v_j).
  \label{corrected}
\end{equation}
In (\ref{corrected}), the left hand expression is valid only in the
the special Cartesian basis normal to the filter box faces, while the
right hand expression is valid generally and is equivalent to the left
hand expression in this special basis.

We point this generalization out here for two reasons. First is that
a valid model must be tensorially consistent, and so it is important
that there is a tensorially consistent expression of the AMD
model. Second, using the tensorially consistent version of the AMD
model allows it to be applied in a broader set of circumstances.
The generalized AMD model expression for the eddy viscosity $\nu_e$
(equation 23 in \cite{roze:2015}) is
given by
\begin{equation}
 \mathcal{R}_{ij}=(\cM_{km}\partial_m v_i)(\cM_{kn}\partial_n
    v_j),
  \label{Rij}
\end{equation}
\begin{equation}
  \nu_e=C\frac{\max(-\mathcal{R}_{ij}S_{ij},0)}{(\partial_k v_j)(\partial_k v_j)},
  \label{AMDgen}
\end{equation}
which is obtained by pushing the generalize scaled gradient through
the development that leads to equation (23) in \cite{roze:2015}.


\begin{acknowledgments}
The authors acknowledge the generous financial support from the
National Aeronautics and Space Administration (cooperative agreement
number NNX15AU40A), the Air Force Office of Scientific Research (grant
FA9550-11-1-007) and the Exascale Computing Project (17-SC-20-SC), a
collaborative effort of two U.S. Department of Energy organizations
(Office of Science and the National Nuclear Security
Administration). Thanks are also due the Texas Advanced Computing
Center at The University of Texas at Austin for providing HPC
resources that have contributed to the research results reported here
(\url{http://www.tacc.utexas.edu}).
\end{acknowledgments}

\bibliography{main}

\begin{thebibliography}{19}%
\makeatletter
\providecommand \@ifxundefined [1]{%
 \@ifx{#1\undefined}
}%
\providecommand \@ifnum [1]{%
 \ifnum #1\expandafter \@firstoftwo
 \else \expandafter \@secondoftwo
 \fi
}%
\providecommand \@ifx [1]{%
 \ifx #1\expandafter \@firstoftwo
 \else \expandafter \@secondoftwo
 \fi
}%
\providecommand \natexlab [1]{#1}%
\providecommand \enquote  [1]{``#1''}%
\providecommand \bibnamefont  [1]{#1}%
\providecommand \bibfnamefont [1]{#1}%
\providecommand \citenamefont [1]{#1}%
\providecommand \href@noop [0]{\@secondoftwo}%
\providecommand \href [0]{\begingroup \@sanitize@url \@href}%
\providecommand \@href[1]{\@@startlink{#1}\@@href}%
\providecommand \@@href[1]{\endgroup#1\@@endlink}%
\providecommand \@sanitize@url [0]{\catcode `\\12\catcode `\$12\catcode
  `\&12\catcode `\#12\catcode `\^12\catcode `\_12\catcode `\%12\relax}%
\providecommand \@@startlink[1]{}%
\providecommand \@@endlink[0]{}%
\providecommand \url  [0]{\begingroup\@sanitize@url \@url }%
\providecommand \@url [1]{\endgroup\@href {#1}{\urlprefix }}%
\providecommand \urlprefix  [0]{URL }%
\providecommand \Eprint [0]{\href }%
\providecommand \doibase [0]{https://doi.org/}%
\providecommand \selectlanguage [0]{\@gobble}%
\providecommand \bibinfo  [0]{\@secondoftwo}%
\providecommand \bibfield  [0]{\@secondoftwo}%
\providecommand \translation [1]{[#1]}%
\providecommand \BibitemOpen [0]{}%
\providecommand \bibitemStop [0]{}%
\providecommand \bibitemNoStop [0]{.\EOS\space}%
\providecommand \EOS [0]{\spacefactor3000\relax}%
\providecommand \BibitemShut  [1]{\csname bibitem#1\endcsname}%
\let\auto@bib@innerbib\@empty
\bibitem [{\citenamefont {Germano}\ \emph {et~al.}(1991)\citenamefont
  {Germano}, \citenamefont {Piomelli}, \citenamefont {Moin},\ and\
  \citenamefont {Cabot}}]{germ:1991}%
  \BibitemOpen
  \bibfield  {author} {\bibinfo {author} {\bibfnamefont {M.}~\bibnamefont
  {Germano}}, \bibinfo {author} {\bibfnamefont {U.}~\bibnamefont {Piomelli}},
  \bibinfo {author} {\bibfnamefont {P.}~\bibnamefont {Moin}},\ and\ \bibinfo
  {author} {\bibfnamefont {W.~H.}\ \bibnamefont {Cabot}},\ }\bibfield  {title}
  {\bibinfo {title} {A dynamic subgrid-scale eddy viscosity model},\
  }\href@noop {} {\bibfield  {journal} {\bibinfo  {journal} {Physics of Fluids
  A: Fluid Dynamics}\ }\textbf {\bibinfo {volume} {3}},\ \bibinfo {pages}
  {1760} (\bibinfo {year} {1991})}\BibitemShut {NoStop}%
\bibitem [{\citenamefont {Lilly}(1992)}]{lilly:1992}%
  \BibitemOpen
  \bibfield  {author} {\bibinfo {author} {\bibfnamefont {D.~K.}\ \bibnamefont
  {Lilly}},\ }\bibfield  {title} {\bibinfo {title} {A proposed modification of
  the germano subgrid scale closure method},\ }\href@noop {} {\bibfield
  {journal} {\bibinfo  {journal} {Physics of Fluids A: Fluid Dynamics}\
  }\textbf {\bibinfo {volume} {4}},\ \bibinfo {pages} {633} (\bibinfo {year}
  {1992})}\BibitemShut {NoStop}%
\bibitem [{\citenamefont {Piomelli}\ and\ \citenamefont
  {Balaras}(2002)}]{piom:2002}%
  \BibitemOpen
  \bibfield  {author} {\bibinfo {author} {\bibfnamefont {U.}~\bibnamefont
  {Piomelli}}\ and\ \bibinfo {author} {\bibfnamefont {E.}~\bibnamefont
  {Balaras}},\ }\bibfield  {title} {\bibinfo {title} {Wall-layer models for
  large-eddy simulations},\ }\href@noop {} {\bibfield  {journal} {\bibinfo
  {journal} {Annual Review of Fluid Mechanics}\ }\textbf {\bibinfo {volume}
  {34}},\ \bibinfo {pages} {349} (\bibinfo {year} {2002})}\BibitemShut
  {NoStop}%
\bibitem [{\citenamefont {Larsson}\ \emph {et~al.}(2016)\citenamefont
  {Larsson}, \citenamefont {Kawai}, \citenamefont {Bodart},\ and\ \citenamefont
  {Bermejo-Moreno}}]{lars:2015}%
  \BibitemOpen
  \bibfield  {author} {\bibinfo {author} {\bibfnamefont {J.}~\bibnamefont
  {Larsson}}, \bibinfo {author} {\bibfnamefont {S.}~\bibnamefont {Kawai}},
  \bibinfo {author} {\bibfnamefont {J.}~\bibnamefont {Bodart}},\ and\ \bibinfo
  {author} {\bibfnamefont {I.}~\bibnamefont {Bermejo-Moreno}},\ }\bibfield
  {title} {\bibinfo {title} {Large eddy simulation with modeled wall-stress:
  recent progress and future directions},\ }\href@noop {} {\bibfield  {journal}
  {\bibinfo  {journal} {Bulletin of JSME}\ }\textbf {\bibinfo {volume} {3}}
  (\bibinfo {year} {2016})}\BibitemShut {NoStop}%
\bibitem [{\citenamefont {Bose}\ and\ \citenamefont {Park}(2018)}]{bose:2018}%
  \BibitemOpen
  \bibfield  {author} {\bibinfo {author} {\bibfnamefont {S.~T.}\ \bibnamefont
  {Bose}}\ and\ \bibinfo {author} {\bibfnamefont {G.~I.}\ \bibnamefont
  {Park}},\ }\bibfield  {title} {\bibinfo {title} {Wall-modeled large-eddy
  simulation for complex turbulent flows},\ }\href@noop {} {\bibfield
  {journal} {\bibinfo  {journal} {Annual Review of Fluid Mechanics}\ }\textbf
  {\bibinfo {volume} {50}},\ \bibinfo {pages} {535} (\bibinfo {year}
  {2018})}\BibitemShut {NoStop}%
\bibitem [{\citenamefont {Scotti}\ \emph {et~al.}(1993)\citenamefont {Scotti},
  \citenamefont {Meneveau},\ and\ \citenamefont {Lilly}}]{scott:1993}%
  \BibitemOpen
  \bibfield  {author} {\bibinfo {author} {\bibfnamefont {A.}~\bibnamefont
  {Scotti}}, \bibinfo {author} {\bibfnamefont {C.}~\bibnamefont {Meneveau}},\
  and\ \bibinfo {author} {\bibfnamefont {D.~K.}\ \bibnamefont {Lilly}},\
  }\bibfield  {title} {\bibinfo {title} {Generalized {S}magorinsky model for
  anisotropic grids},\ }\href@noop {} {\bibfield  {journal} {\bibinfo
  {journal} {Physics of Fluids A}\ }\textbf {\bibinfo {volume} {5}},\ \bibinfo
  {pages} {2306} (\bibinfo {year} {1993})}\BibitemShut {NoStop}%
\bibitem [{\citenamefont {Vreman}(2004)}]{vrem:2004}%
  \BibitemOpen
  \bibfield  {author} {\bibinfo {author} {\bibfnamefont {A.}~\bibnamefont
  {Vreman}},\ }\bibfield  {title} {\bibinfo {title} {An eddy-viscosity
  subgrid-scale model for turbulent shear flow},\ }\href@noop {} {\bibfield
  {journal} {\bibinfo  {journal} {Physics of Fluids}\ }\textbf {\bibinfo
  {volume} {16}},\ \bibinfo {pages} {3670} (\bibinfo {year}
  {2004})}\BibitemShut {NoStop}%
\bibitem [{\citenamefont {Rozema}\ \emph {et~al.}(2015)\citenamefont {Rozema},
  \citenamefont {Bae}, \citenamefont {Moin},\ and\ \citenamefont
  {Verstappen}}]{roze:2015}%
  \BibitemOpen
  \bibfield  {author} {\bibinfo {author} {\bibfnamefont {W.}~\bibnamefont
  {Rozema}}, \bibinfo {author} {\bibfnamefont {H.~J.}\ \bibnamefont {Bae}},
  \bibinfo {author} {\bibfnamefont {P.}~\bibnamefont {Moin}},\ and\ \bibinfo
  {author} {\bibfnamefont {R.}~\bibnamefont {Verstappen}},\ }\bibfield  {title}
  {\bibinfo {title} {Minimum-dissipation models for large-eddy simulation},\
  }\href@noop {} {\bibfield  {journal} {\bibinfo  {journal} {Physics of
  Fluids}\ }\textbf {\bibinfo {volume} {27}},\ \bibinfo {pages} {085107}
  (\bibinfo {year} {2015})}\BibitemShut {NoStop}%
\bibitem [{\citenamefont {Verstappen}(2011)}]{vers:2011}%
  \BibitemOpen
  \bibfield  {author} {\bibinfo {author} {\bibfnamefont {R.}~\bibnamefont
  {Verstappen}},\ }\bibfield  {title} {\bibinfo {title} {When does eddy
  viscosity damp subfilter scales sufficiently?},\ }\href@noop {} {\bibfield
  {journal} {\bibinfo  {journal} {J. Sci. Comput.}\ }\textbf {\bibinfo {volume}
  {49}},\ \bibinfo {pages} {94} (\bibinfo {year} {2011})}\BibitemShut {NoStop}%
\bibitem [{\citenamefont {Jim\'{e}nez}\ and\ \citenamefont
  {Moser}(2000)}]{moser:2000}%
  \BibitemOpen
  \bibfield  {author} {\bibinfo {author} {\bibfnamefont {J.}~\bibnamefont
  {Jim\'{e}nez}}\ and\ \bibinfo {author} {\bibfnamefont {R.~D.}\ \bibnamefont
  {Moser}},\ }\bibfield  {title} {\bibinfo {title} {Large-eddy simulations:
  Where are we and what can we expect?},\ }\href@noop {} {\bibfield  {journal}
  {\bibinfo  {journal} {AIAA Journal}\ }\textbf {\bibinfo {volume} {28}},\
  \bibinfo {pages} {605} (\bibinfo {year} {2000})}\BibitemShut {NoStop}%
\bibitem [{\citenamefont {Smagorinsky}(1963)}]{smag:1963}%
  \BibitemOpen
  \bibfield  {author} {\bibinfo {author} {\bibfnamefont {J.}~\bibnamefont
  {Smagorinsky}},\ }\bibfield  {title} {\bibinfo {title} {General circulation
  experiments with the primitive equations. i. the basic experiment},\
  }\href@noop {} {\bibfield  {journal} {\bibinfo  {journal} {Mon. Weather
  Rev.}\ }\textbf {\bibinfo {volume} {91}},\ \bibinfo {pages} {99} (\bibinfo
  {year} {1963})}\BibitemShut {NoStop}%
\bibitem [{\citenamefont {Lee}\ and\ \citenamefont
  {Moser}(2015)}]{LeeMoser2015}%
  \BibitemOpen
  \bibfield  {author} {\bibinfo {author} {\bibfnamefont {M.}~\bibnamefont
  {Lee}}\ and\ \bibinfo {author} {\bibfnamefont {R.~D.}\ \bibnamefont
  {Moser}},\ }\bibfield  {title} {\bibinfo {title} {Direct numerical simulation
  of turbulent channel flow up to {R}e=5200},\ }\href
  {https://doi.org/10.1017/jfm.2015.268} {\bibfield  {journal} {\bibinfo
  {journal} {Journal of Fluid Mechanics}\ }\textbf {\bibinfo {volume} {774}},\
  \bibinfo {pages} {395} (\bibinfo {year} {2015})}\BibitemShut {NoStop}%
\bibitem [{\citenamefont {Scotti}\ \emph {et~al.}(1996)\citenamefont {Scotti},
  \citenamefont {Meneveau},\ and\ \citenamefont {Fatica}}]{scot:1996b}%
  \BibitemOpen
  \bibfield  {author} {\bibinfo {author} {\bibfnamefont {A.}~\bibnamefont
  {Scotti}}, \bibinfo {author} {\bibfnamefont {C.}~\bibnamefont {Meneveau}},\
  and\ \bibinfo {author} {\bibfnamefont {M.}~\bibnamefont {Fatica}},\
  }\bibfield  {title} {\bibinfo {title} {Dynamic smagorinsky model on
  anisotropic grids},\ }\href@noop {} {\bibfield  {journal} {\bibinfo
  {journal} {Physics of Fluids}\ }\textbf {\bibinfo {volume} {9}} (\bibinfo
  {year} {1996})}\BibitemShut {NoStop}%
\bibitem [{\citenamefont {Synge}\ and\ \citenamefont
  {Schild}(1949)}]{syng:1949}%
  \BibitemOpen
  \bibfield  {author} {\bibinfo {author} {\bibfnamefont {J.~L.}\ \bibnamefont
  {Synge}}\ and\ \bibinfo {author} {\bibfnamefont {A.}~\bibnamefont {Schild}},\
  }\href@noop {} {\emph {\bibinfo {title} {Tensor Calculus}}}\ (\bibinfo
  {publisher} {Dover},\ \bibinfo {year} {1949})\ pp.\ \bibinfo {pages}
  {26--30}\BibitemShut {NoStop}%
\bibitem [{\citenamefont {Chang}\ and\ \citenamefont
  {Moser}(2007)}]{chang:2007}%
  \BibitemOpen
  \bibfield  {author} {\bibinfo {author} {\bibfnamefont {H.}~\bibnamefont
  {Chang}}\ and\ \bibinfo {author} {\bibfnamefont {R.~D.}\ \bibnamefont
  {Moser}},\ }\bibfield  {title} {\bibinfo {title} {An inertial range model for
  the three-point third-order velocity correlation},\ }\href@noop {} {\bibfield
   {journal} {\bibinfo  {journal} {Physics of Fluids}\ }\textbf {\bibinfo
  {volume} {19}} (\bibinfo {year} {2007})}\BibitemShut {NoStop}%
\bibitem [{\citenamefont {Donzis}\ and\ \citenamefont
  {Sreenivasan}(2010)}]{DonzisSreeni2010}%
  \BibitemOpen
  \bibfield  {author} {\bibinfo {author} {\bibfnamefont {D.~A.}\ \bibnamefont
  {Donzis}}\ and\ \bibinfo {author} {\bibfnamefont {K.~R.}\ \bibnamefont
  {Sreenivasan}},\ }\bibfield  {title} {\bibinfo {title} {The bottleneck effect
  and the {K}olmogorov constant in isotropic turbulence},\ }\href@noop {}
  {\bibfield  {journal} {\bibinfo  {journal} {J. Fluid Mech.}\ }\textbf
  {\bibinfo {volume} {657}},\ \bibinfo {pages} {171} (\bibinfo {year}
  {2010})}\BibitemShut {NoStop}%
\bibitem [{\citenamefont {Uribe}\ \emph {et~al.}(2010)\citenamefont {Uribe},
  \citenamefont {Jarrin}, \citenamefont {Prosser},\ and\ \citenamefont
  {Laurence}}]{urib:2010}%
  \BibitemOpen
  \bibfield  {author} {\bibinfo {author} {\bibfnamefont {J.~C.}\ \bibnamefont
  {Uribe}}, \bibinfo {author} {\bibfnamefont {N.}~\bibnamefont {Jarrin}},
  \bibinfo {author} {\bibfnamefont {R.}~\bibnamefont {Prosser}},\ and\ \bibinfo
  {author} {\bibfnamefont {D.}~\bibnamefont {Laurence}},\ }\bibfield  {title}
  {\bibinfo {title} {Development of a two-velocities hybrid {RANS}-{LES} model
  and its application to a trailing edge flow},\ }\href@noop {} {\bibfield
  {journal} {\bibinfo  {journal} {Flow Turbulence Combust}\ }\textbf {\bibinfo
  {volume} {85}},\ \bibinfo {pages} {181?197} (\bibinfo {year}
  {2010})}\BibitemShut {NoStop}%
\bibitem [{\citenamefont {Lilly}(1967)}]{Lilly:1967}%
  \BibitemOpen
  \bibfield  {author} {\bibinfo {author} {\bibfnamefont {D.~K.}\ \bibnamefont
  {Lilly}},\ }\bibfield  {title} {\bibinfo {title} {The representation of small
  scale turbulence in numerical simulation experiments},\ }\href@noop {}
  {\bibfield  {journal} {\bibinfo  {journal} {IBM Scientific Computing
  Symposium on environmental sciences}\ ,\ \bibinfo {pages} {195}} (\bibinfo
  {year} {1967})}\BibitemShut {NoStop}%
\bibitem [{\citenamefont {Haering}\ \emph {et~al.}(2018)\citenamefont
  {Haering}, \citenamefont {Oliver},\ and\ \citenamefont {Moser}}]{haer:2018}%
  \BibitemOpen
  \bibfield  {author} {\bibinfo {author} {\bibfnamefont {S.}~\bibnamefont
  {Haering}}, \bibinfo {author} {\bibfnamefont {T.}~\bibnamefont {Oliver}},\
  and\ \bibinfo {author} {\bibfnamefont {R.~D.}\ \bibnamefont {Moser}},\
  }\bibfield  {title} {\bibinfo {title} {Towards a predictive hybrid
  {RANS}/{LES} framework},\ }\href@noop {} {\bibfield  {journal} {\bibinfo
  {journal} {AIAA Scitech 2019 Forum}\ }\textbf {\bibinfo {volume} {AIAA
  2019-0087}} (\bibinfo {year} {2018})}\BibitemShut {NoStop}%
\end{thebibliography}%

\end{document}